\documentclass[fleqn,usenatbib]{mnras}

\usepackage{newtxtext,newtxmath}
\usepackage[T1]{fontenc}
\usepackage{ae,aecompl}

\usepackage[usenames,dvipsnames]{xcolor}
\usepackage{graphicx}
\usepackage{amsmath}
\usepackage{amssymb}
\usepackage{siunitx}
\usepackage{microtype}

\graphicspath{{./}{figures/}}  

\DeclareSIUnit\MHz{\mega\hertz}
\DeclareSIUnit\kHz{\kilo\hertz}
\DeclareSIUnit\jansky{Jy}
\DeclareSIUnit\mJy{\milli\jansky}
\DeclareSIUnit\mK{\milli\kelvin}
\DeclareSIUnit\parsec{pc}
\DeclareSIUnit\arcsec{arcsec}
\DeclareSIUnit\Mpc{\mega\parsec}

\newcommand{\R}[1]{\mathrm{#1}}
\newcommand{\B}[1]{\mathbfit{#1}}

\title[EoR signal separation with a CDAE]{%
  Separating the EoR signal with a convolutional denoising autoencoder:
  a deep-learning-based method
}

\author[Li~et~al.]{%
Weitian Li,$^{1}$\thanks{E-mail:
  \href{mailto:liweitianux@sjtu.edu.cn}{liweitianux@sjtu.edu.cn} (WL);
  \href{mailto:hgxu@sjtu.edu.cn}{hgxu@sjtu.edu.cn} (HX)}
Haiguang Xu,$^{1,2}$\footnotemark[1]
Zhixian Ma,$^{3}$
Ruimin Zhu,$^{4}$
Dan Hu,$^{1}$
Zhenghao Zhu,$^{1}$
\newauthor  
Junhua Gu,$^{5}$
Chenxi Shan,$^{1}$
Jie Zhu$^{3}$
and
Xiang-Ping Wu$^{5}$
\\
$^{1}${School of Physics and Astronomy,
  Shanghai Jiao Tong University,
  800 Dongchuan Road, Shanghai 200240, China} \\
$^{2}${Tsung-Dao Lee Institute / IFSA Collaborative Innovation Center,
  Shanghai Jiao Tong University,
  800 Dongchuan Road, Shanghai 200240, China} \\
$^{3}${Department of Electronic Engineering,
  Shanghai Jiao Tong University,
  800 Dongchuan Road, Shanghai 200240, China} \\
$^{4}${Department of Statistics,
  Northwestern University,
  2006 Sheridan Road, Evanston, IL 60208, US} \\
$^{5}${National Astronomical Observatories,
  Chinese Academy of Sciences,
  20A Datun Road, Beijing 100012, China}
}

\date{%
  Accepted 2019 February 23.
  Received 2019 February 20;
  in original form 2018 October 14
}

\pubyear{2019}

\begin{document}
\label{firstpage}
\pagerange{\pageref{firstpage}--\pageref{lastpage}}
\maketitle

%
%
\begin{abstract}
When applying the foreground removal methods to uncover the
faint cosmological signal from the epoch of reionization (EoR),
the foreground spectra are assumed to be smooth.
However, this assumption can be seriously violated in practice since
the unresolved or mis-subtracted foreground sources, which are further
complicated by the frequency-dependent beam effects of interferometers,
will generate significant fluctuations along the frequency dimension.
To address this issue, we propose a novel deep-learning-based method
that uses a nine-layer convolutional denoising autoencoder (CDAE) to
separate the EoR signal.
After being trained on the SKA images simulated with realistic beam
effects, the CDAE achieves excellent performance as the mean correlation
coefficient ($\bar{\rho}$) between the reconstructed and input EoR
signals reaches \num{0.929 +- 0.045}.
In comparison,
the two representative traditional methods, namely the polynomial
fitting method and the continuous wavelet transform method, both have
difficulties in modelling and removing the foreground emission
complicated with the beam effects,
yielding only
$\bar{\rho}_{\R{poly}} = \num{0.296 +- 0.121}$ and
$\bar{\rho}_{\R{cwt}} = \num{0.198 +- 0.160}$, respectively.
We conclude that, by hierarchically learning sophisticated features
through multiple convolutional layers, the CDAE is a powerful tool that
can be used to overcome the complicated beam effects and accurately
separate the EoR signal.
Our results also exhibit the great
potential of deep-learning-based methods in future EoR experiments.
\end{abstract}



\section{Introduction}
\label{sec:intro}

The \SI{21}{\cm} line emission of neutral hydrogen from the epoch of
reionization (EoR) is regarded as a decisive probe to directly explore
this stage (see \citealt{furlanetto2016rev} for a review).
To detect the \SI{21}{\cm} signal, which is believed to have been
redshifted to the frequencies below \SI{200}{\MHz}, low-frequency
radio interferometers such as the SKA \citep{koopmans2015rev} and its
pathfinders and precursors have been built or under construction.
The observational challenges, however, are immense due to
complicated instrumental effects, ionospheric distortions, radio frequency
interference, and the strong foreground contamination that
overwhelms the EoR signal by about \numrange{4}{5} orders of magnitude
(see \citealt{morales2010rev} for a review).
Fortunately, in the frequency dimension the foreground contamination
is expected to be intrinsically smooth, while the EoR signal fluctuates
rapidly on \si{\MHz} scales.
This difference is the key characteristic exploited by many foreground
removal methods in order to uncover the faint EoR signal
\citep[e.g.,][]{wang2006,jelic2008,harker2009,liu2009fgrm,%
  chapman2012,chapman2013,gu2013,wang2013,bonaldi2015,mertens2018}.

However, the smoothness of the foreground spectra can be damaged
by the frequency-dependent beam effects, i.e., the variation of the point
spread function (PSF) with frequencies that cannot be perfectly
calibrated \citep{liu2009ps}.
Because of the incomplete $uv$ coverage,
the PSF has a complicated profile consisting of a narrow peaky main lobe
and a multitude of jagged side lobes with relative amplitudes of about
0.1 per cent that extend beyond the field of view
\citep[e.g.,][their figs 1 and 3]{liu2009ps}.
A source that is unresolved or mis-subtracted (e.g., due to the limited
field of view) during the CLEAN process leaves catastrophic residuals,
the locations of which vary with the frequency since the angular
position of a PSF side lobe is inversely proportional to the frequency.
These effects lead to complicated residuals fluctuating along the
frequency dimension, which cannot be correctly separated from the EoR
signal by the traditional foreground removal methods that rely on
the smoothness of the foreground spectra.

Given the complicated profiles and frequency-dependent variations of
the PSF, it would be very difficult to craft a practicable model for most,
if not all, existing foreground removal methods to overcome the beam
effects, even at the cost of extensive computation burden
\citep[e.g.,][]{lochner2015}.
Therefore, deep-learning-based methods, which can distil knowledge from
the data to automatically refine the model, seem more feasible
and appealing \citep[e.g.,][]{herbel2018,vafaeiSadr2019}.
In recent years, deep learning algorithms have seen prosperous
developments and have brought breakthroughs into many fields
(see \citealt{lecun2015} for a recent review).
Among various deep learning algorithms, the autoencoder is a common type of
neural networks that aims at learning useful features from the input data
in an unsupervised manner, and it is usually used for dimensionality
reduction \citep[e.g.,][]{hinton2006,wang2014} and data denoising
\citep[e.g.,][]{xie2012,bengio2013,lu2013}.
In particular,
the convolutional denoising autoencoder (CDAE) is very flexible and
powerful in capturing subtle and complicated features in the data and have
been successfully applied to weak gravitational wave signal denoising
\citep[e.g.,][]{shen2017}, monaural audio source separation
\citep[e.g.,][]{grais2017}, and so on.
These applications have demonstrated the outstanding abilities of the
CDAE in extracting weak signals from highly temporal-variable data.
Thus, it is worth trying to apply the CDAE to separate the EoR signal.
Although the signal-to-noise ratio in the EoR separation task is much lower
than in existing applications, the EoR signal and foreground emission as
well as the beam effects are stationary or semistationary.

In this paper, a novel deep-learning-based method that uses a CDAE
is proposed to tackle the complicated frequency-dependent beam effects
and to separate the EoR signal along the frequency dimension.
In \autoref{sec:method}, we introduce the CDAE and elaborate
the proposed method.
In \autoref{sec:experiments}, we demonstrate the performance of the
CDAE by applying it to the simulated SKA images.
We discuss the method and carry out
comparisons to traditional methods
in \autoref{sec:discussions}.
Finally, we summarize our work in \autoref{sec:summary}.
The implementation code and data are made public at
\url{https://github.com/liweitianux/cdae-eor}.

\section{Methodology}
\label{sec:method}

\subsection{Convolutional denoising autoencoder}
\label{sec:cdae}

An autoencoder is composed of an encoder and a decoder, which can be
characterized by the functions $f(\cdot)$ and $g(\cdot)$, respectively.
The encoder maps the input $\B{x}$ to an internal code $\B{h}$, i.e.,
$\B{h} = f(\B{x})$, and the decoder tries to reconstruct the desired
signal from the code $\B{h}$, i.e., $\B{r} = g(\B{h})$, where $\B{x}$,
$\B{h}$, and $\B{r}$ are all vectors in this work.
By placing constraints (e.g., dimensionality, sparsity) on the
code $\B{h}$ and training the autoencoder to minimize the
loss $L(\B{r}, \,\B{x})$, which quantifies the difference between the
reconstruction $\B{r}$ and the input $\B{x}$, the autoencoder is expected
to learn the codes that effectively represent the input
\citep[chapter 14]{goodfellow2016}.

In order to make the autoencoder learn a better representation of the input
to achieve better performance, \citet{vincent2008,vincent2010} proposed a
novel training strategy based on the denoising criterion:
artificially corrupt the original input $\B{x}$ (e.g., by adding noise),
feed the corrupted input $\B{x}'$ into the autoencoder, and then train it
to reconstruct the original input $\B{x}$ by minimizing the loss
$L(\B{r}, \,\B{x})$.
During this denoising process, the autoencoder is
forced to capture robust features that are critical to accurately
reconstruct the original input.
An autoencoder trained with such a strategy
is called a `denoising autoencoder.'

Classic autoencoders are built with fully connected layers, each neuron of
which is connected with every neuron in the previous layer.
This makes the total number of parameters increase exponentially with the
number of layers.
Meanwhile, the extracted features are forced to be global, which
is suboptimal to represent the input \citep[e.g.,][]{masci2011}.
On the other hand, convolutional layers, as used in convolutional neural
networks (CNNs),
make use of a set of small filters and share their weights among
all locations in the data \citep[e.g.,][]{lecun1998}, which allows to
better capture the local features in the data.
Therefore, CNNs generally have 2 or more orders of magnitude less
parameters than the analogous fully connected neural networks
\citep[e.g.,][]{grais2017}
and require much less training resources such as memory and time.
Furthermore, multiple convolutional layers can be easily stacked to extract
sophisticated higher level features by composing the lower-level ones
obtained in previous layers.
This technique guarantees highly expressive CNNs that achieve
outstanding performance in image classification and relevant fields
\citep[e.g.,][]{krizhevsky2012,simonyan2014,szegedy2015,ma2019}.
By utilising multiple convolutional layers instead of fully connected
layers in a denoising autoencoder, the obtained CDAE gains the powerful
feature extraction capability of CNNs, which helps improve its denoising
performance, and can reconstruct even seriously corrupted signals
\citep[e.g.,][]{du2017}.
In consequence, the CDAE may be well suited to exploit the complicated
differences between the EoR signal and the foreground emission
for the purpose of separating them accurately.

\subsection{Network architecture}
\label{sec:architecture}

\begin{figure*}
  \centering
  \includegraphics[width=0.9\textwidth]{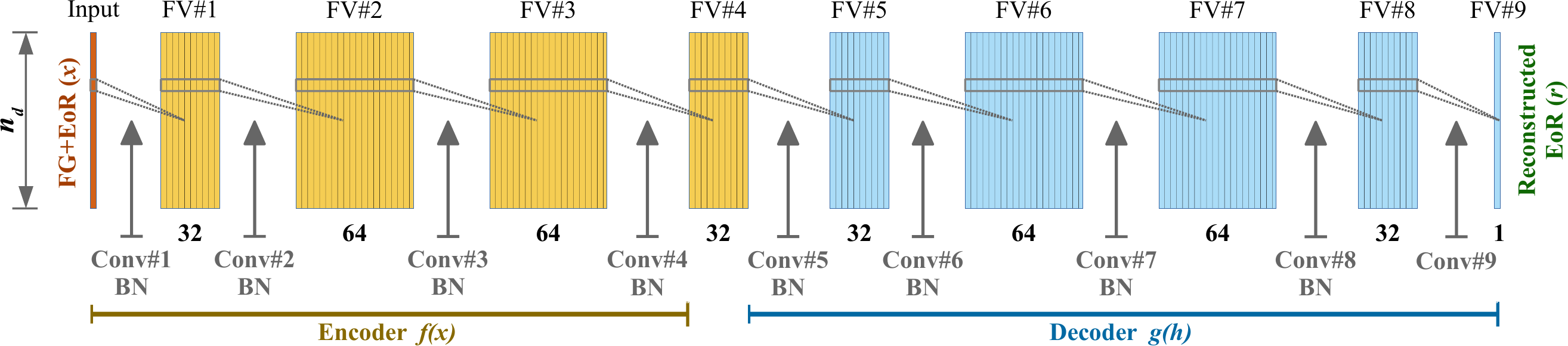}
  \caption{\label{fig:network}%
    The architecture of the proposed CDAE that consists of a four-layer
    encoder and a five-layer decoder.
    The orange and blue boxes represent the feature vectors (FV) generated
    by the encoder and decoder layers, respectively.
    The numbers marked below the boxes are the number of filters in the
    corresponding convolutional layers.
    The batch normalization (BN) technique is applied to all layers except
    for the last layer.
  }
\end{figure*}

Both the encoder and decoder parts of the proposed CDAE consist of multiple
convolutional layers.
We do not set a strict boundary between the two parts
because we focus on the feature extraction and denoising capabilities
rather than on the specific formats of the internal codes $\B{h}$.
For the $l$-th convolutional layer that has $m_l$ filters, a set of $m_l$
feature vectors
$\left(\left\{ \B{v}_{i}^{(l)} \right\}; i = 1, 2, \cdots, m_l \right)$
are generated as the output of this layer by convolving the output of the
previous layer
$\left(\left\{ \B{v}_{j}^{(l-1)} \right\}; j = 1, 2, \cdots, m_{l-1} \right)$
with each of the filters, i.e.,
\begin{equation}
  \label{eq:conv}
  \B{v}_{i}^{(l)} = \phi^{(l)} \left( \sum_{j=1}^{m_{l-1}}
    \B{v}_{j}^{(l-1)} * W_i^{(l)} + b_i^{(l)} \right),
    \quad i = 1, 2, \cdots, m_{l},
\end{equation}
where
$W_i^{(l)}$ and $b_i^{(l)}$ are the weights and bias of this filter
in the $l$-th layer, $\phi^{(l)}(\cdot)$ is the layer's activation
function, and `$*$' denotes the convolution operation.

Following the common practices \citep[e.g.,][]{geron2017,suganuma2018},
we adopt filters of size three in all layers and use the exponential linear
unit (ELU; \citealt{clevert2016}) as the activation function
$\phi^{(l)}(\cdot)$ for all layers except the last layer, which uses
the hyperbolic tangent function (i.e., tanh;
see also \autoref{sec:preprocessing}).
In addition, the batch normalization technique is applied to all layers
except for the last layer to improve the training process as well as to act
as a regularizer to prevent overfitting \citep{ioffe2015}.

To determine the number of convolutional layers and the number of filters
in each layer, we have tested multiple CDAE architectures, each containing
a different number of layers and filters.
After evaluating their performances (see also \autoref{sec:results}),
the simplest one with sufficiently good performance is selected,
which consists of a four-layer encoder with $(32,64,64,32)$ filters and
a five-layer decoder with $(32,64,64,32,1)$ filters, as illustrated in
\autoref{fig:network}.
We note that the pooling and upsampling layers are not used in the CDAE
because they have very little impact on the performance according to our
tests \citep[see also][]{springenberg2015}.

\subsection{Training and evaluation}
\label{sec:train-eval}

At the beginning, the parameters of the CDAE (i.e., the weights and biases
of filters in all layers) are initialized randomly using the He uniform
initializer \citep{he2015}.
In order to obtain an effective CDAE by training these parameters, the
following three data sets are required \citep[e.g.,][]{ripley1996}:
(1) training set ($S_{\R{tr}}$);
(2) validation set ($S_{\R{val}}$) that is used to validate the training
process and to help determine the hyperparameters (e.g., the number of
layers and filters);
(3) test set ($S_{\R{test}}$) that is solely used to evaluate the
performance of the trained CDAE.
Each data set is a collection of many data points of
($\B{x}, \B{x}_{\R{eor}}$),
where $\B{x} = \B{x}_{\R{eor}} + \B{x}_{\R{fg}}$ is the total emission of
one sky pixel, and $\B{x}_{\R{eor}}$ is the corresponding EoR signal.

During each training epoch, the total emission $\B{x}^{(i)}$ is fed into
the CDAE and goes through all the convolutional layers (\autoref{eq:conv}),
outputting the reconstructed EoR signal $\B{r}^{(i)}_{\R{eor}}$.
The difference between the reconstructed EoR signal $\B{r}^{(i)}_{\R{eor}}$
and the input EoR signal $\B{x}^{(i)}_{\R{eor}}$ is the loss $L$ and can be
quantified with the mean squared error (MSE), i.e.,
\begin{equation}
  \label{eq:loss}
  L = \frac{1}{N_{\R{tr}}} \sum_{i=1}^{N_{\R{tr}}}
    \left[ \B{r}_{\R{eor}}^{(i)} - \B{x}_{\R{eor}}^{(i)} \right]^T
    \left[ \B{r}_{\R{eor}}^{(i)} - \B{x}_{\R{eor}}^{(i)} \right],
\end{equation}
where $N_{\R{tr}}$ is the number of data points in the training set
$S_{\R{tr}}$.
By applying the back-propagation method
\citep[e.g.,][]{rumelhart1986,lecun1998bp},
the parameters are updated to reduce the loss $L$, so as to
improve the quality of the reconstructed EoR signal.
As the training goes for more epochs, the initially randomized CDAE is
gradually shaped into a network
that learns a better representation of the input and can
reconstruct the EoR signal more accurately.

To evaluate the performance of the trained CDAE,
the Pearson's correlation coefficient
\citep[e.g.,][]{harker2009,chapman2013}
is adopted to measure the similarity between the reconstructed and input
EoR signals:
\begin{equation}
  \label{eq:corrcoef}
  \rho(\B{r}_{\R{eor}}, \B{x}_{\R{eor}})
      = \frac{\sum_{j=1}^{n}(r_{\R{eor},j} - \bar{r}_{\R{eor}})
      (x_{\R{eor},j} - \bar{x}_{\R{eor}})}{
        \sqrt{\sum_{j=1}^{n}(r_{\R{eor},j} - \bar{r}_{\R{eor}})^2
          \sum_{j=1}^{n}(x_{\R{eor},j} - \bar{x}_{\R{eor}})^2}
    },
\end{equation}
where
$\bar{r}_{\R{eor}}$ and $\bar{x}_{\R{eor}}$ represent the mean
values of $\B{r}_{\R{eor}}$ and $\B{x}_{\R{eor}}$, respectively, and $n$
is the length of the signals.
The closer to one the correlation coefficient
$\rho(\B{r}_{\R{eor}}, \B{x}_{\R{eor}})$ is,
the better the achieved performance.

\section{Experiments}
\label{sec:experiments}

\subsection{Simulation of the SKA images}
\label{sec:simulation}

We carry out end-to-end simulations to generate the SKA images to
train the proposed CDAE and evaluate its performance.
A representative frequency band, namely \SIrange{154}{162}{\MHz}, is chosen
as an example \citep[e.g.,][]{datta2010} and is divided into $n_f = 101$
channels with a resolution of \SI{80}{\kHz}.
At each frequency channel, the sky maps of the foreground emission and the
EoR signal are simulated within an area of \SI{10 x 10}{\degree} and are
pixelized into \num{1800 x 1800} with a pixel size of \SI{20}{\arcsec}.

\begin{figure*}
  \centering
  \includegraphics[width=0.8\textwidth]{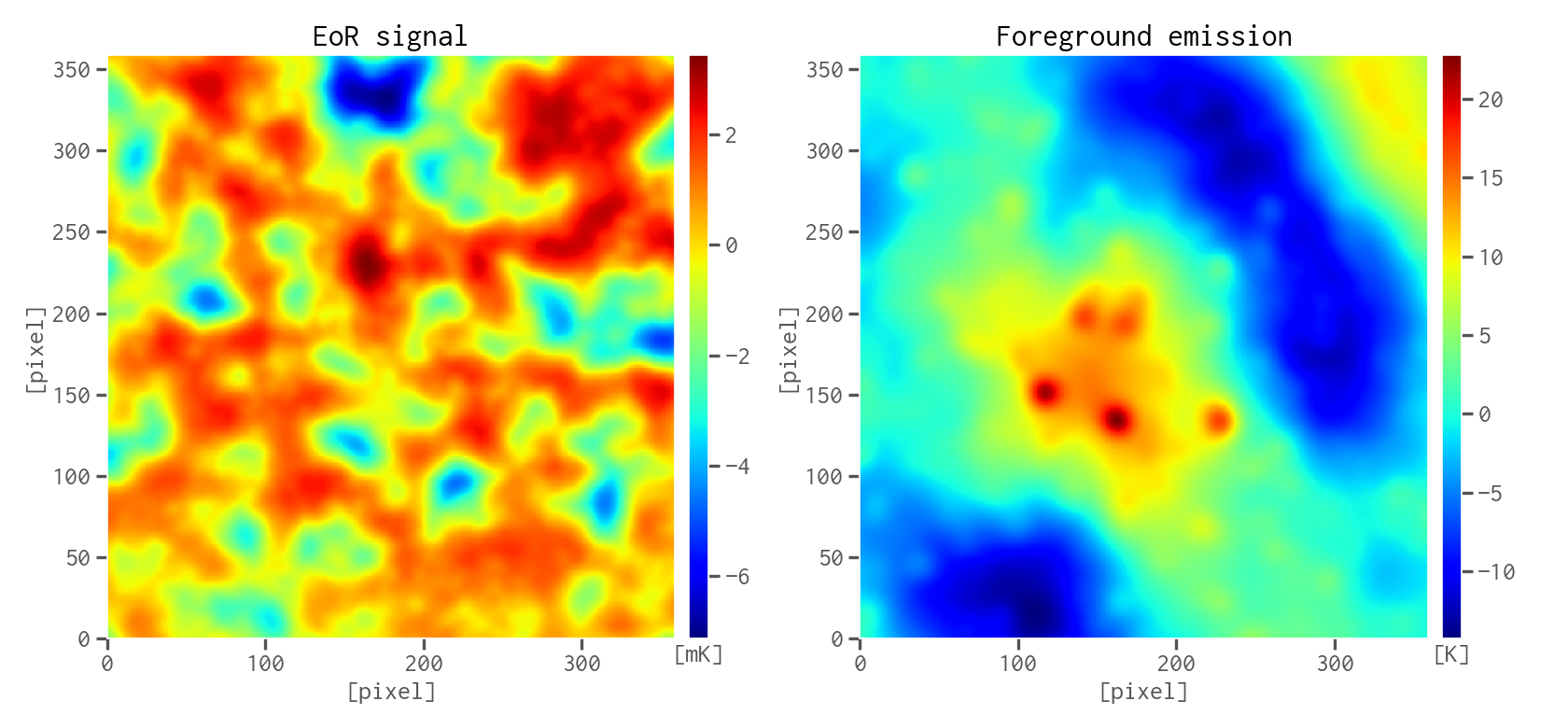}
  \caption{\label{fig:obsimg}%
    Simulated images of the EoR signal (left-hand panel) and the foreground
    emission (right-hand panel) at \SI{158}{\MHz}.
    Both images have sizes of \num{360 x 360} and cover sky areas of
    \SI{2 x 2}{\degree}.
    The blobs in the right-hand panel show the bright point sources and
    radio haloes.
  }
\end{figure*}

Based on our previous work \citep{wang2010}, we simulate the foreground
emission by taking into account the contributions from the Galactic
synchrotron and free-free radiations, extragalactic point sources, and
radio haloes.
The Galactic synchrotron radiation is simulated by extrapolating the Haslam
\SI{408}{\MHz} map with a power-law spectrum, the index of which is given
by the synchrotron spectral index map \citep{giardino2002} to account for
its variation with sky positions.
The reprocessed Haslam \SI{408}{\MHz} map\footnote{%
  The reprocessed Haslam \SI{408}{\MHz} map:
  \url{http://www.jb.man.ac.uk/research/cosmos/haslam_map/}}
\citep{remazeilles2015}, which has significantly better instrument
calibration and more accurate extragalactic source subtraction,
is used as the template to obtain enhanced simulation results over
\citet{wang2010}.
By employing the tight relation between the H$\alpha$ and free-free
emissions \citep[see][and references therein]{dickinson2003}, the Galactic
free-free radiation can be derived from the H$\alpha$ survey map
\citep{finkbeiner2003}.
Since the Galactic diffuse emissions vary remarkably across the sky, we
simulate them at a central position of (R.A., Dec\@.) = (\SI{0}{\degree},
\SI{-27}{\degree}), which has a high galactic latitude
($b = \SI{-78.5}{\degree}$) and is an appropriate choice for the simulation
of SKA images \citep[e.g.,][]{beardsley2016}.
We account for the following five types of extragalactic point sources:
(1) star-forming and starburst galaxies, (2) radio-quiet active galactic
nuclei (AGNs), (3) Fanaroff--Riley type I and type II AGNs, (4) GHz-peaked
spectrum AGNs, and (5) compact steep spectrum AGNs.
The former three types of sources are simulated by utilizing the data
published by \citet{wilman2008} and the latter two types are simulated by
employing their corresponding luminosity functions and spectral models.
Similar to the real-time peeling of the brightest point sources in
practical data analysis pipelines \citep[e.g.,][]{mitchell2008,intema2009},
we assume that sources with a \SI{158}{\MHz} flux density
$S_{158} > \SI{10}{\mJy}$ have been removed \citep[e.g.,][]{liu2009ps}.
The radio haloes are simulated by generating a sample of galaxy clusters
with the Press--Schechter formalism \citep{press1974} and then applying
multiple scaling relations (e.g., between cluster mass and X-ray
temperature, between X-ray temperature and radio power) to derive their
radio emissions.

In regard to the simulation of the EoR signal, we take advantage of the
2016 data release from the
\textit{Evolution Of 21\,cm Structure} project\footnote{%
  Evolution Of 21\,cm Structure:
  \url{http://homepage.sns.it/mesinger/EOS.html}}
\citep{mesinger2016} and extract the image slices at corresponding
redshifts (i.e., frequencies) from the light-cone cube of the recommended
`faint galaxies' case.
The extracted image slices are then re-scaled to match the sky coverage and
pixel size of the foreground maps.

\begin{figure}
  \centering
  \includegraphics[width=\columnwidth]{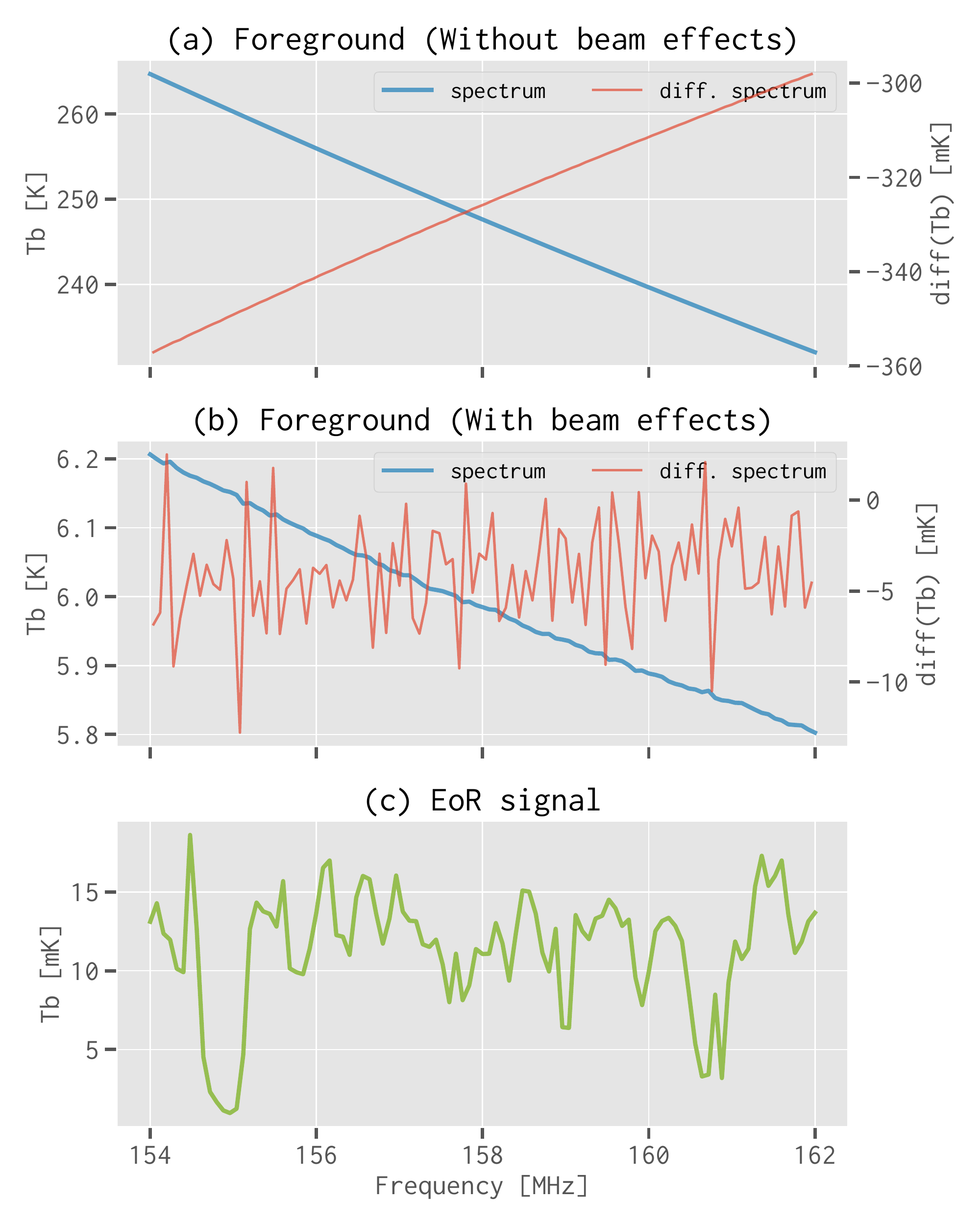}
  \caption{\label{fig:simudata}%
    Example spectra of the foreground emission and the EoR signal for one
    random sky pixel.
    \textbf{Top}: The ideal (i.e., without beam effects) foreground
    spectrum (the blue line) and the corresponding differential spectrum
    (the red line).
    \textbf{Middle}: The `observed' (i.e., with beam effects) foreground
    spectrum (the blue line) and the corresponding differential spectrum
    (the red line).
    \textbf{Bottom}: The EoR signal spectrum (the green line).
  }
\end{figure}

To incorporate the realistic frequency-dependent beam effects into the
simulated sky maps, we further adopt the latest SKA1-Low layout
configuration\footnote{\raggedright%
  SKA1-Low layout:
  \url{https://astronomers.skatelescope.org/wp-content/uploads/2016/09/SKA-TEL-SKO-0000422_02_SKA1_LowConfigurationCoordinates-1.pdf}}
to simulate instrument observations.
The SKA1-Low interferometer is composed of 512 stations, each of which
contains 256 antennas randomly distributed inside a circle of
\SI{35}{\meter} in diameter.
The 512 stations are divided into two parts:
(1) 224 stations are randomly distributed within the `core' region of
\SI{1000}{\meter} in diameter;
(2) the remaining stations are placed on three spiral arms extending up to a
radius of about \SI{35}{\kilo\meter}.
For each sky map, we employ the \textsc{OSKAR}\footnote{%
  OSKAR: \url{https://github.com/OxfordSKA/OSKAR} (version 2.7.0)}
simulator \citep{mort2010} to perform 6-hour synthesis imaging
to obtain the visibility data, from which the `observed'
image is created by the \textsc{WSClean}\footnote{%
  WSClean: \url{https://sourceforge.net/p/wsclean} (version 2.5)}
imager \citep{offringa2014}.
In order to emphasize the faint and relatively diffuse EoR signal, the
natural weighting and baselines of \numrange{30}{1000} wavelengths are
utilized in the imaging process.
Finally, the created images are cropped to keep only the central
\SI{2 x 2}{\degree} regions (i.e., \num{360 x 360} pixels) for the
purpose of the best quality.
Therefore, we obtain a pair of image cubes of size
\num{360 x 360 x 101} for the EoR signal $\left( C_{\R{eor}}^{(1)} \right)$
and the foreground emission $\left( C_{\R{fg}}^{(1)} \right)$, respectively
(see \autoref{fig:obsimg} for the simulated images at the central frequency
of \SI{158}{\MHz}).
To better illustrate the impacts of beam effects on the foreground spectra,
we take one random sky pixel as an example and show the foreground
spectra with and without the beam effects in \autoref{fig:simudata}, where
the corresponding differential spectra (i.e., differences between every
two adjacent frequency channels) and the EoR signal spectrum are also
plotted.
Compared to the ideal sky foreground (the top panel), the spectral
smoothness of the `observed' foreground (the middle panel) is seriously
damaged by the rapid fluctuations resulted from the beam effects.
Although such fluctuations exhibit somewhat similar spectral scales
($< \SI{1}{\MHz}$) as the EoR signal (the bottom panel), they are
still sufficiently different, which can be exploited by the CDAE to achieve
an effective separation.
We note that the `observed' foreground has an amplitude of about two orders
of magnitude smaller than the ideal foreground, the major reason for which
is that interferometers are only sensitive to the spatial fluctuations of
the emission \citep[e.g.,][]{braun1985}.

Considering that the training and evaluation of the CDAE require three
data sets (i.e., training, validation, and test; \autoref{sec:train-eval}),
if there are only one pair of image cubes, the test set $S_{\R{test}}$
could only contain a small fraction of all the pixels that are randomly
distributed on the sky, from which it is impossible to obtain a complete
image of the reconstructed EoR signal.
Consequently, it is beneficial to simulate another pair of image cubes
that are solely used as the test set.
To this end, we simulate the Galactic diffuse radiations at a central
coordinate of (R.A., Dec\@.) = (\SI{3}{\degree}, \SI{-27}{\degree}), i.e.,
\SI{3}{\degree} away from the first pair of image cubes, which is
sufficient because the finally cropped image cubes only cover a sky area of
\SI{2 x 2}{\degree}.
Since extragalactic point sources, radio haloes, and the EoR signal are
mostly isotropic, we shift their sky maps simulated above by
\SI{3}{\degree} to generate the new sky maps.
Following the same procedures to simulate instrument observations, we
obtain the second pair of image cubes
$\left( C_{\R{eor}}^{(2)}, C_{\R{fg}}^{(2)} \right)$.

We note that the simulations do not include thermal noise because the
proposed method is designed to create tomographic EoR images from very deep
observations that have a sufficiently low noise level.
The SKA1-Low is planned to observe each of the target fields for
about \SI{1000}{\hour}, reaching an unprecedented image noise level of
$\,\lesssim \SI{1}{\mK}$ that allows to directly image the reionization
structures \citep[e.g.,][]{mellema2013rev,mellema2015,koopmans2015rev}.

\subsection{Data pre-processing}
\label{sec:preprocessing}

The data set $S = \{(\B{x}, \,\B{x}_{\R{eor}})\}$ for the CDAE is derived
from the simulated image cubes $C_{\R{eor}}$ and $C_{\R{fg}}$, each data
point $(\B{x} = \B{x}_{\R{eor}} + \B{x}_{\R{fg}}, \,\B{x}_{\R{eor}})$
representing the total emission and the EoR signal of one sky pixel,
respectively.
The data set thus has $N_S = \num{360x360 x 2} = \num{259200}$
data points in total.

For the input data $X = \{\B{x}\}$, we propose to apply the
Fourier Transform (FT) along the frequency dimension,
which makes the EoR signal more distinguishable from the
foreground emission and thus easier to be learned by the CDAE
(a comparison with the results derived without applying the FT is
presented in \autoref{sec:why-ft}).
The Blackman--Nuttall window function is applied to suppress the
FT side lobes caused by the sharp discontinuities at both ends
of the finite frequency band \citep[e.g.,][]{chapman2016}.
It is sufficient to keep only half the Fourier coefficients because
$\B{x}$ is real, thus $\B{x}$ of length $n_f = 101$ is transformed to
be 51 complex Fourier coefficients.
The $n_{\R{ex}}$ coefficients of the lowest Fourier frequencies are
excised since they are mostly contributed by the spectral-smooth
foreground emission.
We adopt $n_{\R{ex}} = 6$ to achieve a balance between the
foreground emission suppression and the EoR signal loss.
The real and imaginary parts of the remaining 45 complex coefficients
are then concatenated into a new real vector of length $n_d = 90$,
since the CDAE requires real data.
Finally, the data are zero-centred and normalized to have unit variance.

The pre-processing steps for the input EoR signal
$X_{\R{eor}} = \{\B{x}_{\R{eor}}\}$
are basically the same except for minor adjustments.
After applying the FT, excising the $n_{\R{ex}}$ lowest Fourier
components, and concatenating the real and imaginary parts,
the data elements that have a value less than the 1st
percentile or greater than the 99th percentile are truncated,
in order to prevent the possible outliers from hindering the training of
the CDAE.
Finally, the value range of the data is scaled to $[-1, 1]$ by
dividing by the maximum absolute value,
which allows to use the `tanh' activation function whose value range
is also $[-1, 1]$ in the output layer of the proposed CDAE
(\autoref{sec:architecture}).

\subsection{Training and results}
\label{sec:results}

The pre-processed data of the first cube pair
$\left( C_{\R{eor}}^{(1)}, C_{\R{fg}}^{(1)} \right)$
are randomly partitioned into the training set ($S_{\R{tr}}$; corresponding
to 80 per cent of the pixels, or \num{103680} data points) and the
validation set ($S_{\R{val}}$; 20 per cent, or \num{25920} data points).
The pre-processed data of the second cube pair
$\left( C_{\R{eor}}^{(2)}, C_{\R{fg}}^{(2)} \right)$
are solely used as the test set ($S_{\R{test}}$; \num{129600} data points).

We implement the proposed CDAE using the
\textsc{Keras}\footnote{Keras: \url{https://keras.io} (version 2.2.4)}
framework \citep{keras} with the
\textsc{TensorFlow}\footnote{TensorFlow:
  \url{https://www.tensorflow.org} (version 1.12.0)}
back end \citep{tensorflow},
which is accelerated by the \textsc{cuda}\footnote{\raggedright%
  CUDA: \url{https://developer.nvidia.com/cuda-zone} (version 9.1.85)}
toolkit.
We adopt a small initial learning rate ($\alpha = \num{e-5}$) and use the
Adam optimization method \citep{kingma2015}.
The CDAE is trained on the training set ($S_{\R{tr}}$) with a batch size of
100 until the training loss converges, which takes about 50 epochs.

\begin{figure}
  \centering
  \includegraphics[width=\columnwidth]{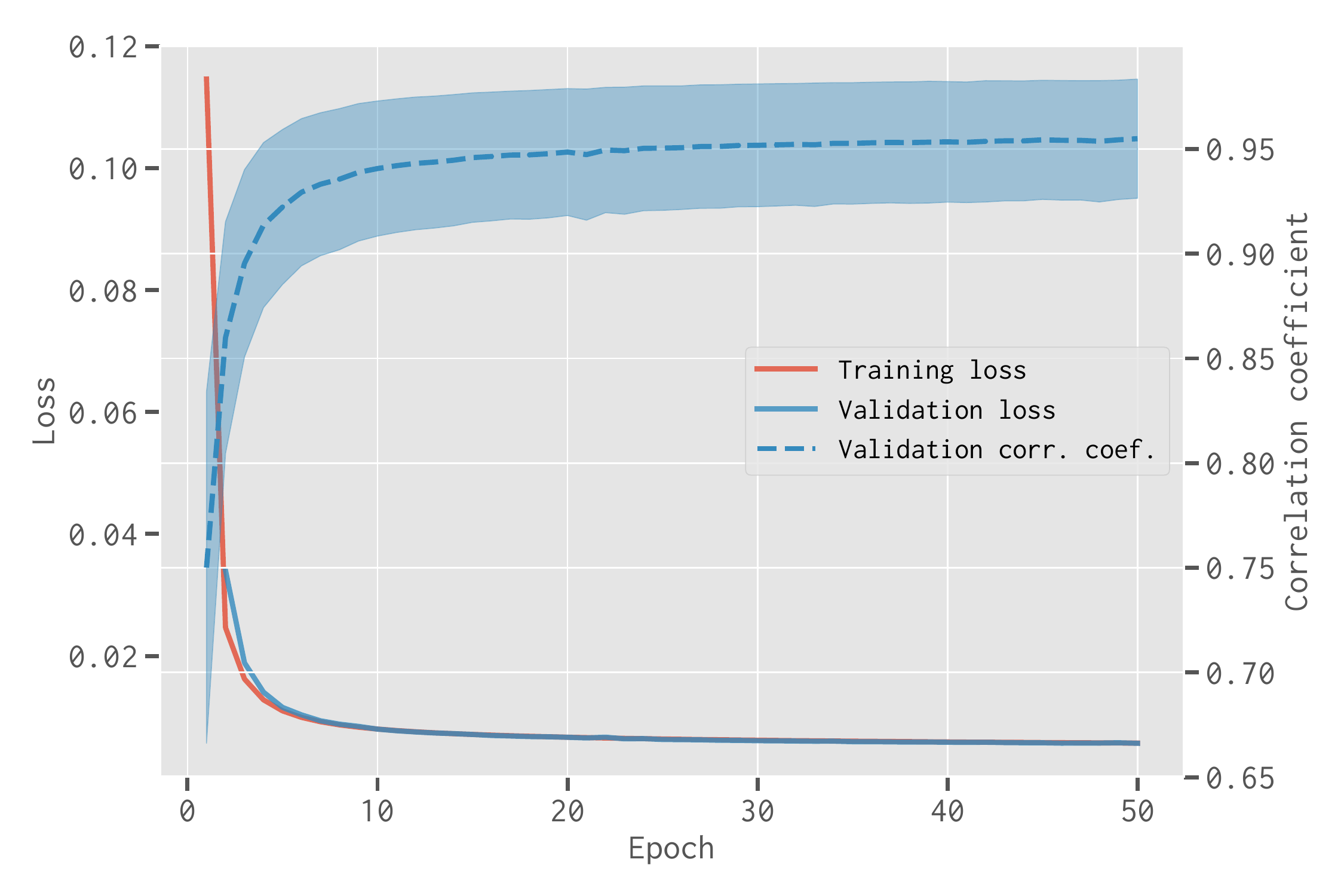}
  \caption{\label{fig:train}%
    The training loss (the solid red line), validation loss (the solid blue
    line), and correlation coefficient ($\rho$; the dashed blue
    line with the shaded region representing its standard deviation)
    calculated on the validation set $S_{\R{val}}$ along the training of
    the CDAE.
  }
\end{figure}

\begin{figure}
  \centering
  \includegraphics[width=\columnwidth]{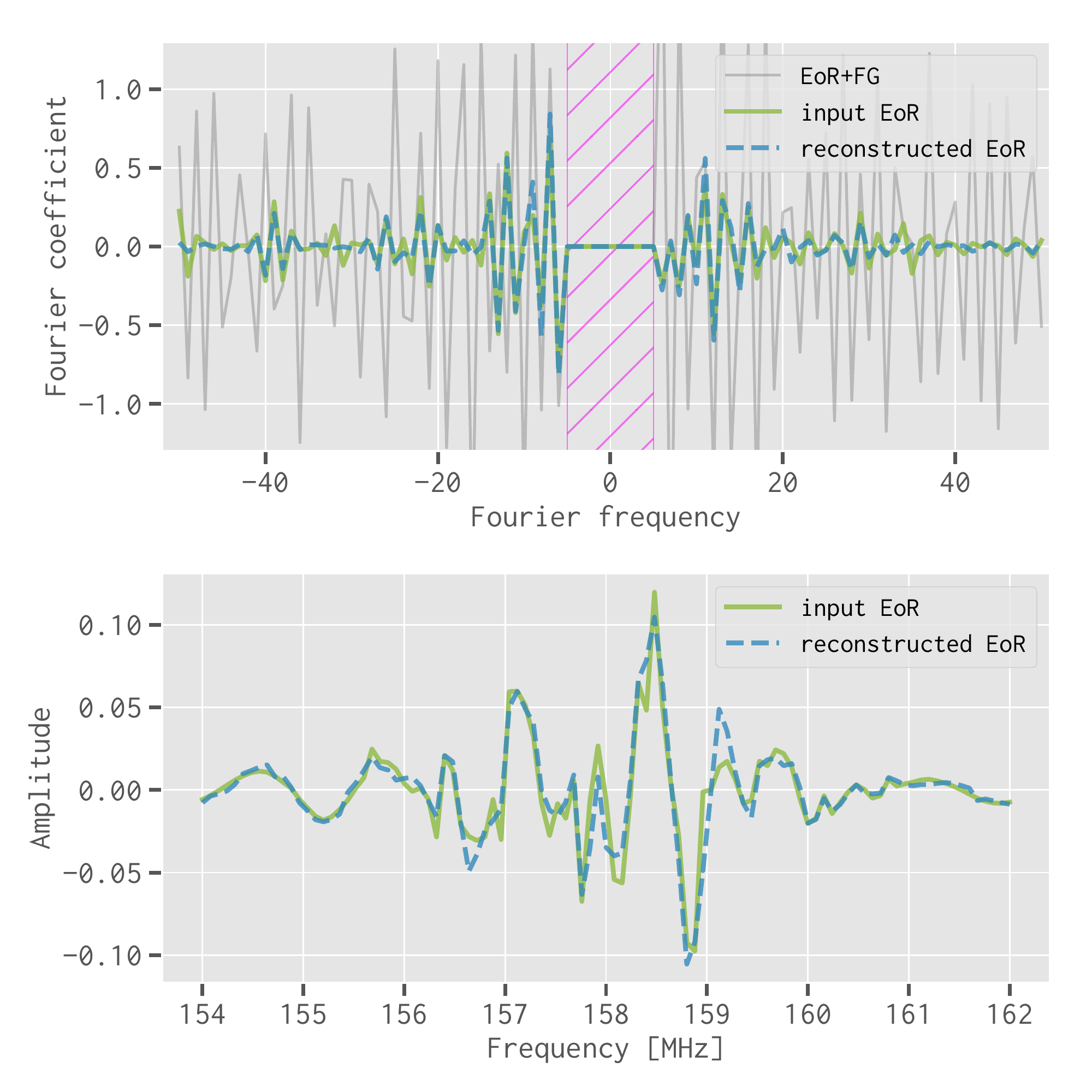}
  \caption{\label{fig:eor-pix}%
    An example of the EoR signal reconstructed by the trained CDAE for
    one pixel in $S_{\R{test}}$.
    \textbf{(Top)} The input EoR signal $\B{x}_{\R{eor}}$ (the solid
    green line) and the reconstructed EoR signal $\B{r}_{\R{eor}}$
    (the dashed blue line) in the Fourier domain.
    The correlation coefficient between the input and reconstructed EoR
    signals is $\rho = 0.931$.
    The grey line represents the input total emission
    $\B{x} = \B{x}_{\R{fg}} + \B{x}_{\R{eor}}$.
    The magenta hatched region marks the excised Fourier coefficients
    in data pre-processing.
    \textbf{(Bottom)} The input EoR signal $\B{x}_{\R{eor}}$ (the solid
    green line) and the reconstructed EoR signal $\B{r}_{\R{eor}}$
    (the dashed blue line) transformed back to the observing frequency
    domain.
  }
\end{figure}

The training and validation losses together with the evaluation index
(i.e., the correlation coefficient $\rho$) calculated on the validation set
$S_{\R{val}}$ during the training phase are shown in \autoref{fig:train}.
The steadily decreasing losses and increasing correlation coefficient
suggest that the CDAE is well trained without over-fitting.
After training, the evaluation with the test set $S_{\R{test}}$ yields a
high correlation coefficient of $\bar{\rho}_{\R{CDAE}} = \num{0.929 +- 0.045}$
between the reconstructed and input EoR signals.
This result demonstrates that the trained CDAE achieves excellent
performance in reconstructing the EoR signal.
As an example, \autoref{fig:eor-pix} illustrates the reconstructed EoR
signal ($\rho = 0.931$) for one pixel in $S_{\R{test}}$.

\begin{figure*}
  \centering
  \includegraphics[width=0.8\textwidth]{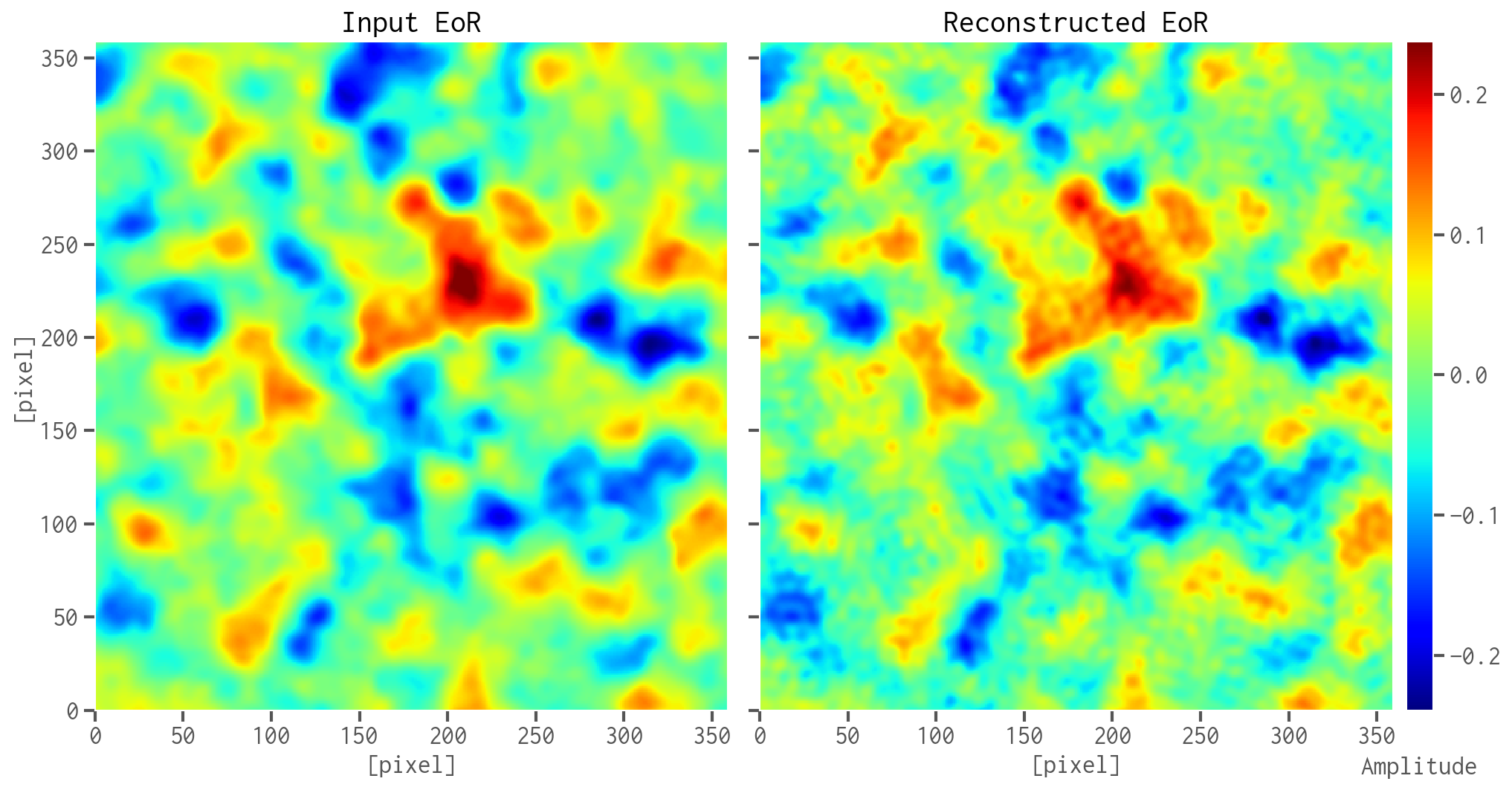}
  \caption{\label{fig:eor-img}%
    Comparison between the input EoR image (left-hand panel) and
    reconstructed EoR image (right-hand panel) at the central frequency of
    \SI{158}{\MHz}.
    The images have the same size (\num{360 x 360} pixel) and the figures
    share the same colour bar (the amplitude is normalized for the CDAE).
  }
\end{figure*}

\begin{figure*}
  \centering
  \includegraphics[width=0.8\textwidth]{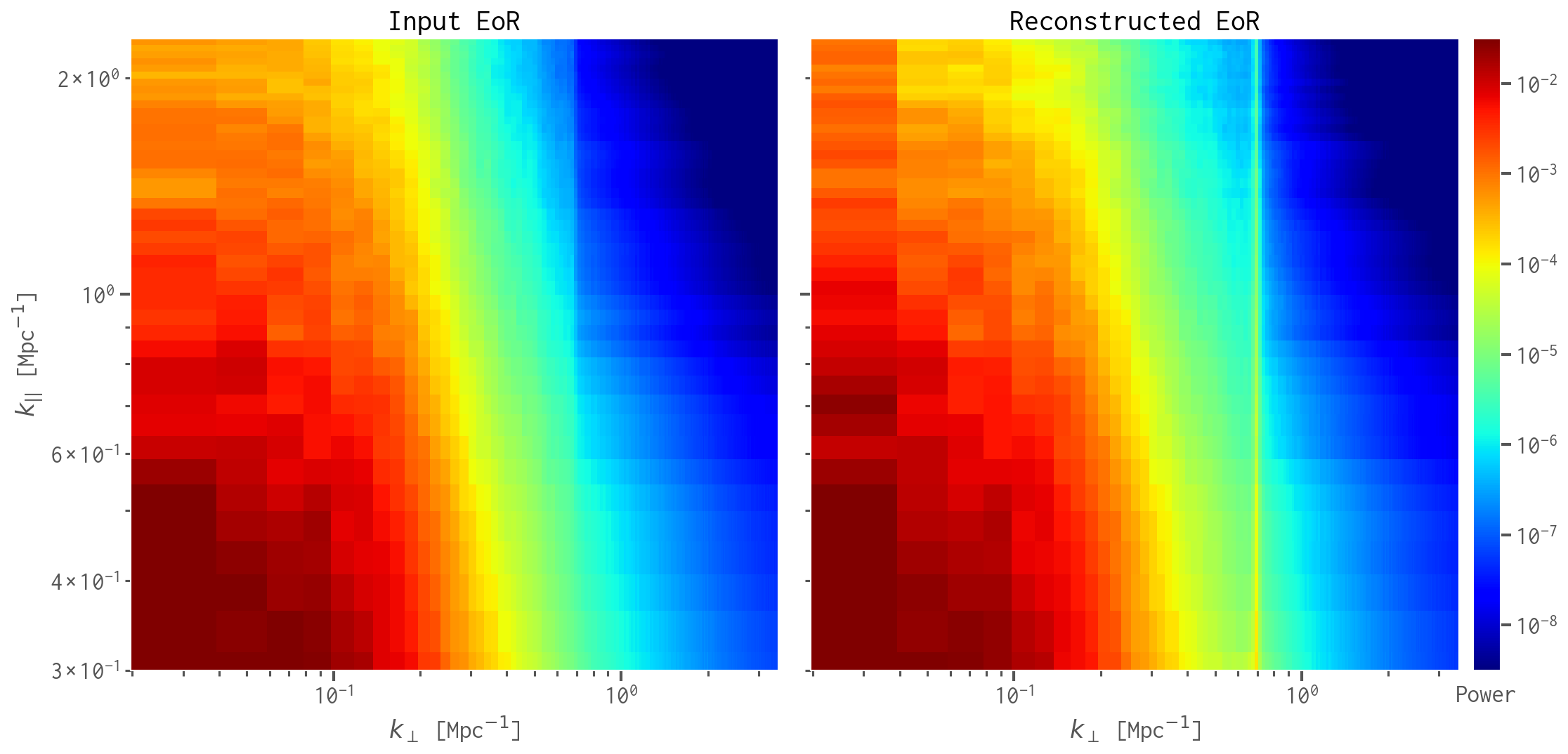}
  \caption{\label{fig:eor-ps}%
    Comparison of two-dimensional power spectra between the input
    (left-hand panel) and reconstructed (right-hand panel) EoR signals.
  }
\end{figure*}

Since the test set $S_{\R{test}}$ is derived from the whole image cubes
$\left( C_{\R{eor}}^{(2)}, C_{\R{fg}}^{(2)} \right)$, we are able to create
complete images of the reconstructed EoR signal and calculate the
corresponding power spectrum.
Taking the input and reconstructed EoR images at the central frequency of
\SI{158}{\MHz} as an example (\autoref{fig:eor-img}), the reconstructed EoR
signal exhibits almost identical structures and amplitudes as the input EoR
signal.
We note that the reconstructed EoR image has weak but detectable redundant
ripples on scales of about 10 pixels (i.e., \SI{200}{\arcsec}),
which are associated with the excision of the $n_{\R{ex}} = 6$ lowest
Fourier frequencies in data pre-processing (\autoref{sec:preprocessing}).
In addition, we calculate the two-dimensional power spectra from the image
cubes of the input and reconstructed EoR signals (\autoref{fig:eor-ps}).
It illustrates that the trained CDAE well recovers the EoR signal on all
covered scales except for a very thin stripe region at
$k_{\bot} \approx \SI{0.7}{\per\Mpc}$, where extra powers are generated
by the aforementioned ripples in the reconstructed EoR images.
We also note that there is a barely visible line at
$k_{\bot} \approx \SI{0.1}{\per\Mpc}$ in both power spectra, which is
caused by the boundary effect of Fourier transforming the finite frequency
band.

The results clearly demonstrate that the trained CDAE is able to accurately
reconstruct the EoR signal, overcoming the complicated beam effects.
The achieved excellent performance of the CDAE can be mainly attributed
to the architecture of stacking multiple convolutional layers, which
implements a powerful feature extraction technique by hierarchically
combining the basic features learned in each layer to build more and
more sophisticated features \citep{lecun2015}.
Combined with the flexibility provided by the \num{53569} trainable
parameters, the CDAE, after being well trained, can intelligently learn a
model that is optimised to accurately separate the faint EoR signal
\citep[e.g.,][]{domingos2012}.

\subsection{Further validation of the CDAE}
\label{sec:validation}

\begin{figure*}
  \centering
  \includegraphics[width=0.8\textwidth]{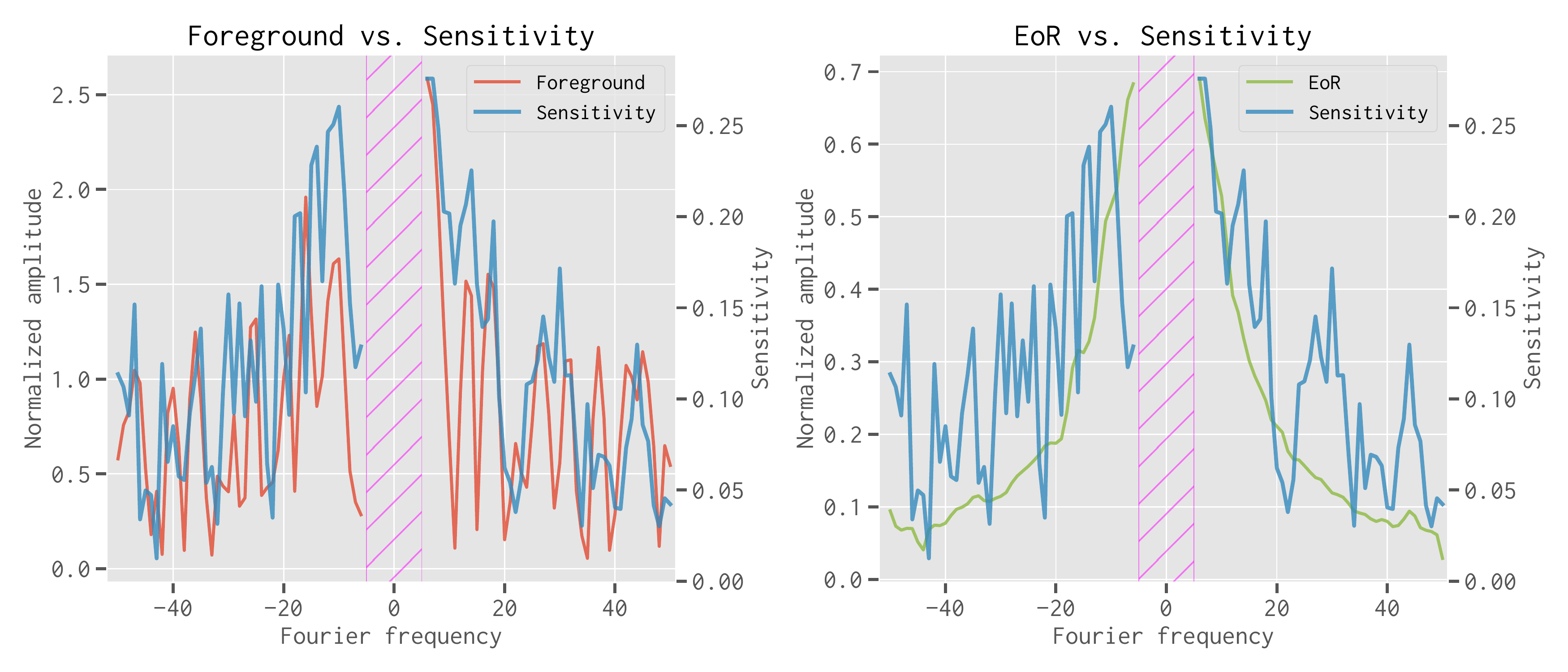}
  \caption{\label{fig:occ-fgeor}%
    The CDAE's sensitivity distribution $\B{s}$ (blue lines in both
    panels) obtained by applying the occlusion method.
    We also plot the root-mean-square amplitudes of the foreground emission
    ($\B{y}_{\R{fg}}$, red line in the left-hand panel) and the EoR signal
    ($\B{y}_{\R{eor}}$, green line in the right-hand panel).
    The sensitivity distribution $\B{s}$ is more correlated with the EoR
    signal [$\rho(\B{s}, \B{y}_{\R{eor}}) = 0.742$] than the foreground
    [$\rho(\B{s}, \B{y}_{\R{fg}}) = 0.562$].
  }
\end{figure*}

With the purpose of further validating that the trained CDAE has actually
learned the useful features of the EoR signal, we employ the occlusion
method \citep{zeiler2014} to visualize the sensitivity of the trained CDAE
to the different part of the input data.
At each time, we occlude three adjacent elements of every input
$\B{x}$ in the validation set $S_{\R{val}}$, and then measure the CDAE's
sensitivity to the occluded part, which is calculated as the
occlusion-induced performance loss, i.e.,
\begin{equation}
  \label{eq:perf-loss}
  s = \frac{1}{N_{\R{val}}} \sum_{i=1}^{N_{\R{val}}}
    \left[ \rho\left(\B{r}^{(i)}_{\R{eor}}, \B{x}^{(i)}_{\R{eor}}\right) -
      \rho\left(\B{R}^{(i)}_{\R{eor}}, \B{x}^{(i)}_{\R{eor}}\right) \right],
\end{equation}
where
$N_{\R{val}}$ is the number of data points in the validation set,
$\B{x}^{(i)}_{\R{eor}}$ is the input EoR signal, and
$\B{r}^{(i)}_{\R{eor}}$ and $\B{R}^{(i)}_{\R{eor}}$ are the reconstructed
EoR signals without and with applying the occlusion, respectively.
By varying the occlusion part of the input data and calculating the
sensitivities, we obtain the CDAE's sensitivity distribution ($\B{s}$) to
every part of the input data, as shown in \autoref{fig:occ-fgeor}, where
the root-mean-square amplitudes of the foreground emission
($\B{y}_{\R{fg}}$) and the EoR signal ($\B{y}_{\R{eor}}$) are also plotted.
We find that the sensitivity distribution is more correlated with the EoR
signal [$\rho(\B{s}, \B{y}_{\R{eor}}) = 0.742$] than the foreground
[$\rho(\B{s}, \B{y}_{\R{fg}}) = 0.562$].
This verifies that the trained CDAE has learned useful features of the EoR
signal to distinguish it from the foreground emission and thus becomes more
sensitive to the data parts of higher signal-to-noise ratio.

\section{Discussions}
\label{sec:discussions}

\subsection{Why pre-process the data set with Fourier Transform?}
\label{sec:why-ft}

\begin{figure}
  \centering
  \includegraphics[width=\columnwidth]{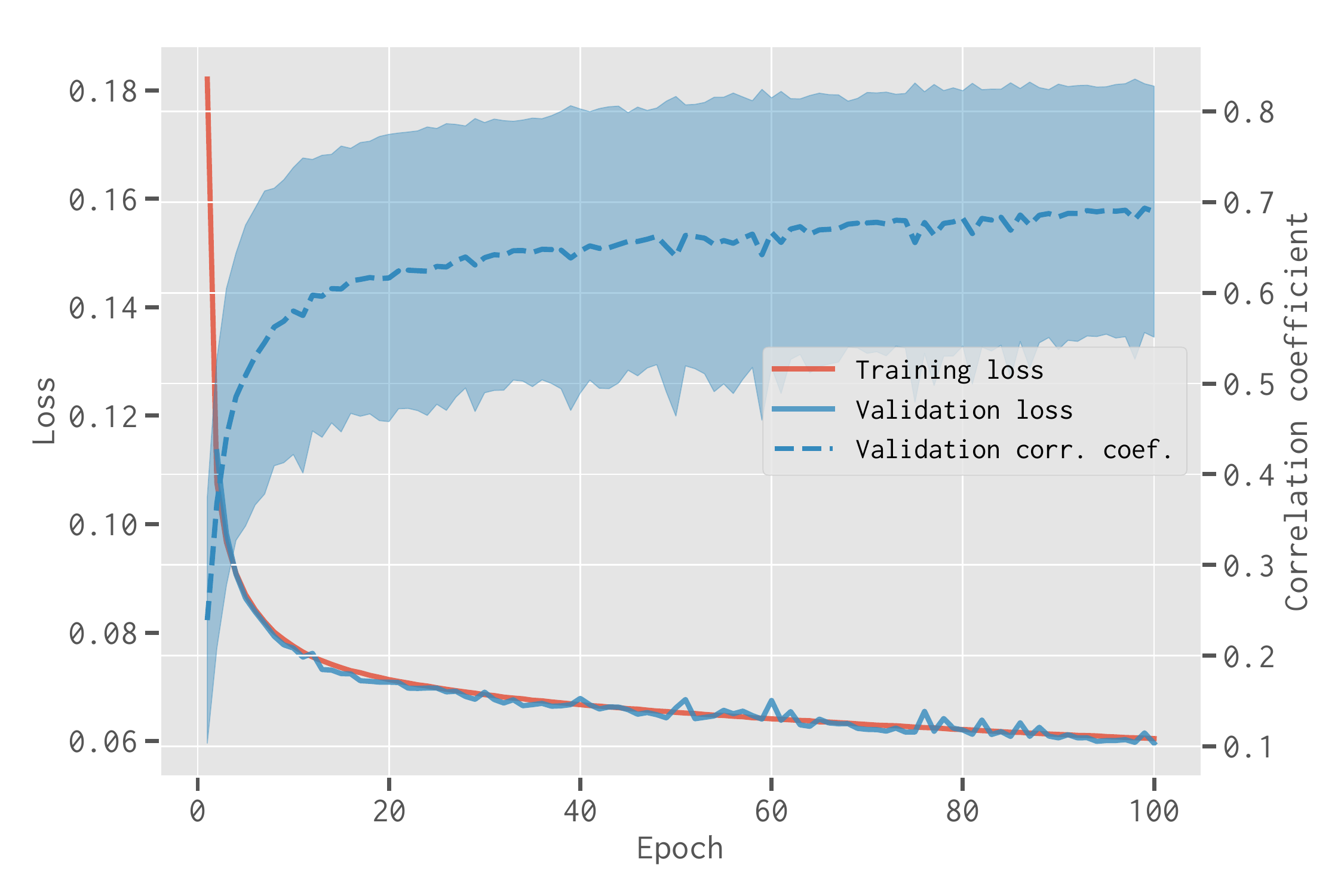}
  \caption{\label{fig:train-noft}%
    Same as \autoref{fig:train} but for the case that the data are
    pre-processed without applying the FT.
  }
\end{figure}

We perform another experiment using the same CDAE architecture,
data sets, and data pre-processing steps, but without
applying the FT as depicted in \autoref{sec:preprocessing}.
After training the CDAE in the same way as described in
\autoref{sec:results}, the correlation coefficient between the
reconstructed and input EoR signals evaluated on the test set
$S_{\R{test}}$ reaches only $\bar{\rho}_{\R{noft}} = \num{0.628 +- 0.167}$,
which indicates a significantly worse performance compared to the case with
FT applied.
As presented in \autoref{fig:train-noft}, the training loss decreases more
slowly and converges after about 100 epochs.
We also find that the training process is slightly unstable given the small
spikes on the curves of both the loss and correlation coefficient.
These indicate that it is beneficial to pre-process the
data set by applying the FT along the frequency dimension, because the
EoR signal and the foreground emission become more distinguishable
in the Fourier domain, where the fluctuating EoR signal concentrates on
larger Fourier modes while the spectral-smooth foreground emission
distributes mainly on smaller Fourier modes \citep[e.g.,][]{parsons2012}.

\subsection{Comparing to traditional methods}
\label{sec:comparisons}

A variety of methods have been proposed to remove the foreground
contamination with the aim of revealing the faint EoR signal.
These methods can be broadly classified into two categories:
(1) parametric methods that apply a parametric model (e.g., a low-degree
polynomial) to fit and remove the foreground emission
\citep[e.g.,][]{wang2006,jelic2008,liu2009fgrm,wang2013,bonaldi2015};
(2) non-parametric methods, which do not assume a specific parametric model
for the foreground emission but exploit the differences between the
foreground emission and the EoR signal (e.g., their different spectral
features) to separate them
\citep[e.g.,][]{harker2009,chapman2012,chapman2013,gu2013,mertens2018}.

In order to further demonstrate the performance of our method, we compare
it to two representative traditional methods:
the polynomial fitting method \citep[e.g.,][]{wang2006} and
the continuous wavelet transform (CWT) method \citep{gu2013}.
The polynomial fitting method is the best representative of the parametric
methods because it is widely used due to its simplicity and robustness
\citep[e.g.,][]{jelic2008,liu2009ps,pritchard2010}
and has been compared to various other foreground removal methods
\citep[e.g.,][]{harker2009,alonso2015,chapman2015}.
Among the non-parametric category, the CWT method is chosen since it
performs similarly well as other non-parametric methods, such as the
Wp smoothing method \citep{harker2009} and the generalized morphological
component analysis method \citep{chapman2013},
meanwhile it is faster and simpler \citep{gu2013,chapman2015}.

With the polynomial fitting method,
a low-degree polynomial is fitted along the frequency dimension for each
sky pixel in the image cube of the total emission (i.e.,
$C_{\R{tot}} = C_{\R{eor}} + C_{\R{fg}}$).
Then by subtracting the fitted smooth component, which is regarded as
the foreground emission, the EoR signal is expected to be uncovered.
Using the same image cubes
$\left( C_{\R{eor}}^{(2)}, C_{\R{fg}}^{(2)} \right)$
simulated in \autoref{sec:simulation},
we have tested polynomials of the degree from 2 (quadratic) to
5 (quintic), and find that the quartic polynomial (degree of 4)
can give the best result.
However, the correlation coefficient calculated for the separated EoR
signal in such a case is only
$\bar{\rho}_{\R{poly}} = \num{0.296 +- 0.121}$,
which indicates that the polynomial fitting method performs poorly in
removing the foreground emission.

The CWT method works based on the same assumption as other foreground
removal methods that the foreground emission is spectrally smooth while the
EoR signal fluctuates rapidly along the frequency dimension.
After applying the CWT, the foreground emission and the EoR signal locate
at different positions in the wavelet space because of their different
spectral scales.
Therefore, the foreground emission can be easily separated from the EoR
signal and be removed \citep{gu2013}.
For each sky pixel, the spectrum of the total emission is transformed into
the wavelet space by applying the CWT with the Morlet wavelet function.
In the wavelet space, after identifying and removing the coefficients that
are mainly contributed by the foreground emission, the remaining
coefficients are transformed back to the frequency space to obtain the
spectrum with the foreground emission removed, which is expected to be the
EoR signal.
By evaluating on the same data set
$\left( C_{\R{eor}}^{(2)}, C_{\R{fg}}^{(2)} \right)$,
we have tuned the method parameters (minimum scale $s_{\R{min}}$, maximum
scale $s_{\R{max}}$, number of scales $n_s$, and cone of influence $c_i$)
and adopt $s_{\R{min}} = 7.4$, $s_{\R{max}} = 50.0$, $n_s = 50$, and
$c_i = 1.6$ to obtain the relatively best performance, which is, however,
only $\bar{\rho}_{\R{cwt}} = \num{0.198 +- 0.160}$.
We note that the CWT method performs slightly worse than the polynomial
fitting method, which is different from the comparison in \citet{gu2013}.
This may be caused by the more serious boundary effect since our simulated
data have a narrower bandwidth and coarser frequency resolution than those
of \citet{gu2013}.

The main reason that both traditional foreground removal methods only
obtain remarkably inferior results is that the smoothness of the foreground
spectra is seriously damaged by the frequency-dependent beam effects, which
cause rapid fluctuations of strength the same order as the EoR signal on
the originally smooth foreground spectra (\autoref{fig:simudata}b).
As a result, the foreground spectra complicated by the beam effects cannot
be well fitted by a low-degree polynomial and have more similar spectral
scales as the EoR signal.
In consequence, both methods are unable to well model the complicated
foreground spectra and thus have great difficulties in removing them.
On the contrary, given its data-driven nature and powerful feature
extraction capabilities, the CDAE is able to distil knowledge from the
training data and learns the features to distinguish the EoR signal from
the fluctuations arising from the beam effects.
Hence, the CDAE achieves superior performance in separating the EoR signal.

\section{Summary}
\label{sec:summary}

The frequency-dependent beam effects of interferometers can cause
rapid fluctuations along the frequency dimension,
which damage the smoothness of the foreground spectra and prevent
traditional foreground removal methods from uncovering the EoR signal.
Given the difficulties in crafting practicable models to overcome the
complicated beam effects, methods that can intelligently learn tailored
models from the data seem more feasible and appealing.
To this end, we have proposed a deep-learning-based method that uses
a nine-layer CDAE to separate the EoR signal.
The CDAE has been trained on the simulated SKA images and has achieved
excellent performance.
We conclude that the CDAE has outstanding ability to overcome the
complicated beam effects and accurately separate the faint EoR signal,
exhibiting the great potential of deep-learning-based methods
to play an important role in the forthcoming EoR experiments.

\section*{Acknowledgements}

We thank the reviewer and editor for their useful comments that
greatly help improve the manuscript.
We also thank Jeffrey Hsu for reading the manuscript and providing
helpful suggestions.
This work is supported by
the Ministry of Science and Technology of China
(grant nos\@. 2018YFA0404601, 2017YFF0210903),
and the National Natural Science Foundation of China
(grant nos\@. 11433002, 11621303, 11835009, 61371147).


\bibliographystyle{mnras}
\bibliography{main}

\begin{thebibliography}{}
\makeatletter
\relax
\def\mn@urlcharsother{\let\do\@makeother \do\$\do\&\do\#\do\^\do\_\do\%\do\~}
\def\mn@doi{\begingroup\mn@urlcharsother \@ifnextchar [ {\mn@doi@}
  {\mn@doi@[]}}
\def\mn@doi@[#1]#2{\def\@tempa{#1}\ifx\@tempa\@empty \href
  {http://dx.doi.org/#2} {doi:#2}\else \href {http://dx.doi.org/#2} {#1}\fi
  \endgroup}
\def\mn@eprint#1#2{\mn@eprint@#1:#2::\@nil}
\def\mn@eprint@arXiv#1{\href {http://arxiv.org/abs/#1} {{\tt arXiv:#1}}}
\def\mn@eprint@dblp#1{\href {http://dblp.uni-trier.de/rec/bibtex/#1.xml}
  {dblp:#1}}
\def\mn@eprint@#1:#2:#3:#4\@nil{\def\@tempa {#1}\def\@tempb {#2}\def\@tempc
  {#3}\ifx \@tempc \@empty \let \@tempc \@tempb \let \@tempb \@tempa \fi \ifx
  \@tempb \@empty \def\@tempb {arXiv}\fi \@ifundefined
  {mn@eprint@\@tempb}{\@tempb:\@tempc}{\expandafter \expandafter \csname
  mn@eprint@\@tempb\endcsname \expandafter{\@tempc}}}

\bibitem[\protect\citeauthoryear{{Abadi} et~al.,}{{Abadi}
  et~al.}{2016}]{tensorflow}
{Abadi} M.,  et~al., 2016, in Proceedings of 12th USENIX Symposium on Operating
  Systems Design and Implementation (OSDI 2016). USENIX Association, \url
  {https://www.tensorflow.org/}

\bibitem[\protect\citeauthoryear{{Alonso}, {Bull}, {Ferreira}  \&
  {Santos}}{{Alonso} et~al.}{2015}]{alonso2015}
{Alonso} D.,  {Bull} P.,  {Ferreira} P.~G.,   {Santos} M.~G.,  2015, \mn@doi
  [\mnras] {10.1093/mnras/stu2474}, \href
  {http://adsabs.harvard.edu/abs/2015MNRAS.447..400A} {447, 400}

\bibitem[\protect\citeauthoryear{{Beardsley} et~al.,}{{Beardsley}
  et~al.}{2016}]{beardsley2016}
{Beardsley} A.~P.,  et~al., 2016, \mn@doi [\apj] {10.3847/1538-4357/833/1/102},
  \href {http://adsabs.harvard.edu/abs/2016ApJ...833..102B} {833, 102}

\bibitem[\protect\citeauthoryear{Bengio, Yao, Alain  \& Vincent}{Bengio
  et~al.}{2013}]{bengio2013}
Bengio Y.,  Yao L.,  Alain G.,   Vincent P.,  2013, in Proceedings of the 26th
  International Conference on Neural Information Processing Systems (NIPS
  2013). Curran Associates Inc., USA, pp 899--907, \url
  {http://dl.acm.org/citation.cfm?id=2999611.2999712}

\bibitem[\protect\citeauthoryear{{Bonaldi} \& {Brown}}{{Bonaldi} \&
  {Brown}}{2015}]{bonaldi2015}
{Bonaldi} A.,  {Brown} M.~L.,  2015, \mn@doi [\mnras] {10.1093/mnras/stu2601},
  \href {http://adsabs.harvard.edu/abs/2015MNRAS.447.1973B} {447, 1973}

\bibitem[\protect\citeauthoryear{{Braun} \& {Walterbos}}{{Braun} \&
  {Walterbos}}{1985}]{braun1985}
{Braun} R.,  {Walterbos} R.~A.~M.,  1985, \aap, \href
  {http://adsabs.harvard.edu/abs/1985A%26A...143..307B} {143, 307}

\bibitem[\protect\citeauthoryear{{Chapman} et~al.,}{{Chapman}
  et~al.}{2012}]{chapman2012}
{Chapman} E.,  et~al., 2012, \mn@doi [\mnras]
  {10.1111/j.1365-2966.2012.21065.x}, \href
  {http://adsabs.harvard.edu/abs/2012MNRAS.423.2518C} {423, 2518}

\bibitem[\protect\citeauthoryear{{Chapman} et~al.,}{{Chapman}
  et~al.}{2013}]{chapman2013}
{Chapman} E.,  et~al., 2013, \mn@doi [\mnras] {10.1093/mnras/sts333}, \href
  {http://adsabs.harvard.edu/abs/2013MNRAS.429..165C} {429, 165}

\bibitem[\protect\citeauthoryear{{Chapman} et~al.,}{{Chapman}
  et~al.}{2015}]{chapman2015}
{Chapman} E.,  et~al., 2015, Advancing Astrophysics with the Square Kilometre
  Array (AASKA14), \href {http://adsabs.harvard.edu/abs/2015aska.confE...5C}
  {p.~5}

\bibitem[\protect\citeauthoryear{{Chapman}, {Zaroubi}, {Abdalla}, {Dulwich},
  {Jeli{\'c}}  \& {Mort}}{{Chapman} et~al.}{2016}]{chapman2016}
{Chapman} E.,  {Zaroubi} S.,  {Abdalla} F.~B.,  {Dulwich} F.,  {Jeli{\'c}} V.,
   {Mort} B.,  2016, \mn@doi [\mnras] {10.1093/mnras/stw161}, \href
  {http://adsabs.harvard.edu/abs/2016MNRAS.458.2928C} {458, 2928}

\bibitem[\protect\citeauthoryear{Chollet et~al.}{Chollet et~al.}{2015}]{keras}
Chollet F.,  et~al., 2015, Keras, \url{https://keras.io}

\bibitem[\protect\citeauthoryear{{Clevert}, {Unterthiner}  \&
  {Hochreiter}}{{Clevert} et~al.}{2016}]{clevert2016}
{Clevert} D.-A.,  {Unterthiner} T.,   {Hochreiter} S.,  2016, in The
  International Conference on Learning Representations (ICLR 2016).
  (\mn@eprint {arXiv} {1511.07289})

\bibitem[\protect\citeauthoryear{{Datta}, {Bowman}  \& {Carilli}}{{Datta}
  et~al.}{2010}]{datta2010}
{Datta} A.,  {Bowman} J.~D.,   {Carilli} C.~L.,  2010, \mn@doi [\apj]
  {10.1088/0004-637X/724/1/526}, \href
  {http://adsabs.harvard.edu/abs/2010ApJ...724..526D} {724, 526}

\bibitem[\protect\citeauthoryear{{Dickinson}, {Davies}  \& {Davis}}{{Dickinson}
  et~al.}{2003}]{dickinson2003}
{Dickinson} C.,  {Davies} R.~D.,   {Davis} R.~J.,  2003, \mn@doi [\mnras]
  {10.1046/j.1365-8711.2003.06439.x}, \href
  {http://adsabs.harvard.edu/abs/2003MNRAS.341..369D} {341, 369}

\bibitem[\protect\citeauthoryear{Domingos}{Domingos}{2012}]{domingos2012}
Domingos P.,  2012, \mn@doi [Communications of the ACM]
  {10.1145/2347736.2347755}, 55, 78

\bibitem[\protect\citeauthoryear{Du, Xiong, Wu, Zhang, Zhang  \& Tao}{Du
  et~al.}{2017}]{du2017}
Du B.,  Xiong W.,  Wu J.,  Zhang L.,  Zhang L.,   Tao D.,  2017, \mn@doi [IEEE
  Transactions on Cybernetics] {10.1109/TCYB.2016.2536638}, 47, 1017

\bibitem[\protect\citeauthoryear{{Finkbeiner}}{{Finkbeiner}}{2003}]{finkbeiner2003}
{Finkbeiner} D.~P.,  2003, \mn@doi [\apjs] {10.1086/374411}, \href
  {http://adsabs.harvard.edu/abs/2003ApJS..146..407F} {146, 407}

\bibitem[\protect\citeauthoryear{{Furlanetto}}{{Furlanetto}}{2016}]{furlanetto2016rev}
{Furlanetto} S.~R.,  2016, \mn@doi [Understanding the Epoch of Cosmic
  Reionization: Challenges and Progress] {10.1007/978-3-319-21957-8_9}, \href
  {http://adsabs.harvard.edu/abs/2016ASSL..423..247F} {423, 247}

\bibitem[\protect\citeauthoryear{G\'eron}{G\'eron}{2017}]{geron2017}
G\'eron A.,  2017, {Hands-On Machine Learning with Scikit-Learn and TensorFlow:
  Concepts, Tools, and Techniques to Build Intelligent Systems}, 1st edn.
O'Reilly Media, Inc.

\bibitem[\protect\citeauthoryear{{Giardino}, {Banday}, {G{\'o}rski}, {Bennett},
  {Jonas}  \& {Tauber}}{{Giardino} et~al.}{2002}]{giardino2002}
{Giardino} G.,  {Banday} A.~J.,  {G{\'o}rski} K.~M.,  {Bennett} K.,  {Jonas}
  J.~L.,   {Tauber} J.,  2002, \mn@doi [\aap] {10.1051/0004-6361:20020285},
  \href {http://adsabs.harvard.edu/abs/2002A%26A...387...82G} {387, 82}

\bibitem[\protect\citeauthoryear{{Goodfellow}, {Bengio}  \&
  {Courville}}{{Goodfellow} et~al.}{2016}]{goodfellow2016}
{Goodfellow} I.,  {Bengio} Y.,   {Courville} A.,  2016, Deep Learning.
MIT Press, \url {http://www.deeplearningbook.org}

\bibitem[\protect\citeauthoryear{{Grais} \& {Plumbley}}{{Grais} \&
  {Plumbley}}{2017}]{grais2017}
{Grais} E.~M.,  {Plumbley} M.~D.,  2017, in 5th IEEE Global Conference on
  Signal and Information Processing (GlobalSIP 2017). IEEE, pp 1265--1269
  (\mn@eprint {arXiv} {1703.08019}), \mn@doi{10.1109/GlobalSIP.2017.8309164}

\bibitem[\protect\citeauthoryear{{Gu}, {Xu}, {Wang}, {An}  \& {Chen}}{{Gu}
  et~al.}{2013}]{gu2013}
{Gu} J.,  {Xu} H.,  {Wang} J.,  {An} T.,   {Chen} W.,  2013, \mn@doi [\apj]
  {10.1088/0004-637X/773/1/38}, \href
  {http://adsabs.harvard.edu/abs/2013ApJ...773...38G} {773, 38}

\bibitem[\protect\citeauthoryear{{Harker} et~al.,}{{Harker}
  et~al.}{2009}]{harker2009}
{Harker} G.,  et~al., 2009, \mn@doi [\mnras]
  {10.1111/j.1365-2966.2009.15081.x}, \href
  {http://adsabs.harvard.edu/abs/2009MNRAS.397.1138H} {397, 1138}

\bibitem[\protect\citeauthoryear{He, Zhang, Ren  \& Sun}{He
  et~al.}{2015}]{he2015}
He K.,  Zhang X.,  Ren S.,   Sun J.,  2015, in Proceedings of the 2015 IEEE
  International Conference on Computer Vision (ICCV 2015). IEEE Computer
  Society, Washington DC, USA, pp 1026--1034, \mn@doi{10.1109/ICCV.2015.123}

\bibitem[\protect\citeauthoryear{{Herbel}, {Kacprzak}, {Amara}, {Refregier}  \&
  {Lucchi}}{{Herbel} et~al.}{2018}]{herbel2018}
{Herbel} J.,  {Kacprzak} T.,  {Amara} A.,  {Refregier} A.,   {Lucchi} A.,
  2018, \mn@doi [Journal of Cosmology and Astro-Particle Physics]
  {10.1088/1475-7516/2018/07/054}, \href
  {https://ui.adsabs.harvard.edu/#abs/2018JCAP...07..054H} {2018, 054}

\bibitem[\protect\citeauthoryear{{Hinton} \& {Salakhutdinov}}{{Hinton} \&
  {Salakhutdinov}}{2006}]{hinton2006}
{Hinton} G.~E.,  {Salakhutdinov} R.~R.,  2006, \mn@doi [Science]
  {10.1126/science.1127647}, \href
  {http://adsabs.harvard.edu/abs/2006Sci...313..504H} {313, 504}

\bibitem[\protect\citeauthoryear{{Intema}, {van der Tol}, {Cotton}, {Cohen},
  {van Bemmel}  \& {R{\"o}ttgering}}{{Intema} et~al.}{2009}]{intema2009}
{Intema} H.~T.,  {van der Tol} S.,  {Cotton} W.~D.,  {Cohen} A.~S.,  {van
  Bemmel} I.~M.,   {R{\"o}ttgering} H.~J.~A.,  2009, \mn@doi [\aap]
  {10.1051/0004-6361/200811094}, \href
  {http://adsabs.harvard.edu/abs/2009A%26A...501.1185I} {501, 1185}

\bibitem[\protect\citeauthoryear{Ioffe \& Szegedy}{Ioffe \&
  Szegedy}{2015}]{ioffe2015}
Ioffe S.,  Szegedy C.,  2015, in Proceedings of the 32nd International
  Conference on International Conference on Machine Learning (ICML 2015). PMLR,
  pp 448--456

\bibitem[\protect\citeauthoryear{{Jeli{\'c}} et~al.,}{{Jeli{\'c}}
  et~al.}{2008}]{jelic2008}
{Jeli{\'c}} V.,  et~al., 2008, \mn@doi [\mnras]
  {10.1111/j.1365-2966.2008.13634.x}, \href
  {http://adsabs.harvard.edu/abs/2008MNRAS.389.1319J} {389, 1319}

\bibitem[\protect\citeauthoryear{{Kingma} \& {Ba}}{{Kingma} \&
  {Ba}}{2015}]{kingma2015}
{Kingma} D.~P.,  {Ba} J.,  2015, in International Conference on Learning
  Representations (ICLR 2015).  (\mn@eprint {arXiv} {1412.6980})

\bibitem[\protect\citeauthoryear{{Koopmans} et~al.,}{{Koopmans}
  et~al.}{2015}]{koopmans2015rev}
{Koopmans} L.,  et~al., 2015, Advancing Astrophysics with the Square Kilometre
  Array (AASKA14), \href {http://adsabs.harvard.edu/abs/2015aska.confE...1K}
  {p.~1}

\bibitem[\protect\citeauthoryear{Krizhevsky, Sutskever  \& Hinton}{Krizhevsky
  et~al.}{2012}]{krizhevsky2012}
Krizhevsky A.,  Sutskever I.,   Hinton G.~E.,  2012, in Proceedings of the 25th
  International Conference on Neural Information Processing Systems (NIPS).
  Curran Associates Inc., USA, pp 1097--1105, \url
  {http://dl.acm.org/citation.cfm?id=2999134.2999257}

\bibitem[\protect\citeauthoryear{LeCun, Bottou, Orr  \& M\"{u}ller}{LeCun
  et~al.}{1998a}]{lecun1998bp}
LeCun Y.,  Bottou L.,  Orr G.~B.,   M\"{u}ller K.-R.,  1998a, in Neural
  Networks: Tricks of the Trade. Springer-Verlag, London, UK, UK, pp 9--50,
  \url {http://dl.acm.org/citation.cfm?id=645754.668382}

\bibitem[\protect\citeauthoryear{LeCun, Bottou, Bengio  \& Haffner}{LeCun
  et~al.}{1998b}]{lecun1998}
LeCun Y.,  Bottou L.,  Bengio Y.,   Haffner P.,  1998b, \mn@doi [Proceedings of
  the IEEE] {10.1109/5.726791}, 86, 2278

\bibitem[\protect\citeauthoryear{{LeCun}, {Bengio}  \& {Hinton}}{{LeCun}
  et~al.}{2015}]{lecun2015}
{LeCun} Y.,  {Bengio} Y.,   {Hinton} G.,  2015, \mn@doi [\nat]
  {10.1038/nature14539}, \href
  {http://adsabs.harvard.edu/abs/2015Natur.521..436L} {521, 436}

\bibitem[\protect\citeauthoryear{{Liu}, {Tegmark}  \& {Zaldarriaga}}{{Liu}
  et~al.}{2009a}]{liu2009ps}
{Liu} A.,  {Tegmark} M.,   {Zaldarriaga} M.,  2009a, \mn@doi [\mnras]
  {10.1111/j.1365-2966.2009.14426.x}, \href
  {http://adsabs.harvard.edu/abs/2009MNRAS.394.1575L} {394, 1575}

\bibitem[\protect\citeauthoryear{{Liu}, {Tegmark}, {Bowman}, {Hewitt}  \&
  {Zaldarriaga}}{{Liu} et~al.}{2009b}]{liu2009fgrm}
{Liu} A.,  {Tegmark} M.,  {Bowman} J.,  {Hewitt} J.,   {Zaldarriaga} M.,
  2009b, \mn@doi [\mnras] {10.1111/j.1365-2966.2009.15156.x}, \href
  {http://adsabs.harvard.edu/abs/2009MNRAS.398..401L} {398, 401}

\bibitem[\protect\citeauthoryear{{Lochner}, {Natarajan}, {Zwart}, {Smirnov},
  {Bassett}, {Oozeer}  \& {Kunz}}{{Lochner} et~al.}{2015}]{lochner2015}
{Lochner} M.,  {Natarajan} I.,  {Zwart} J. T.~L.,  {Smirnov} O.,  {Bassett}
  B.~A.,  {Oozeer} N.,   {Kunz} M.,  2015, \mn@doi [\mnras]
  {10.1093/mnras/stv679}, \href
  {https://ui.adsabs.harvard.edu/#abs/2015MNRAS.450.1308L} {450, 1308}

\bibitem[\protect\citeauthoryear{Lu, Tsao, Matsuda  \& Hori}{Lu
  et~al.}{2013}]{lu2013}
Lu X.,  Tsao Y.,  Matsuda S.,   Hori C.,  2013, in 14th Annual Conference of
  the International Speech Communication Association (INTERSPEECH 2013). pp
  436--440, \url
  {https://www.isca-speech.org/archive/interspeech_2013/i13_0436.html}

\bibitem[\protect\citeauthoryear{{Ma} et~al.,}{{Ma} et~al.}{2019}]{ma2019}
{Ma} Z.,  et~al., 2019, \apjs, \href
  {http://adsabs.harvard.edu/abs/2018arXiv181207190M} {240, 34}

\bibitem[\protect\citeauthoryear{Masci, Meier, Cire\c{s}an  \&
  Schmidhuber}{Masci et~al.}{2011}]{masci2011}
Masci J.,  Meier U.,  Cire\c{s}an D.,   Schmidhuber J.,  2011, in Proceedings
  of the 21th International Conference on Artificial Neural Networks (ICANN
  2011). Springer-Verlag, pp 52--59, \url
  {http://dl.acm.org/citation.cfm?id=2029556.2029563}

\bibitem[\protect\citeauthoryear{{Mellema} et~al.,}{{Mellema}
  et~al.}{2013}]{mellema2013rev}
{Mellema} G.,  et~al., 2013, \mn@doi [Experimental Astronomy]
  {10.1007/s10686-013-9334-5}, \href
  {http://adsabs.harvard.edu/abs/2013ExA....36..235M} {36, 235}

\bibitem[\protect\citeauthoryear{{Mellema}, {Koopmans}, {Shukla}, {Datta},
  {Mesinger}  \& {Majumdar}}{{Mellema} et~al.}{2015}]{mellema2015}
{Mellema} G.,  {Koopmans} L.,  {Shukla} H.,  {Datta} K.~K.,  {Mesinger} A.,
  {Majumdar} S.,  2015, Advancing Astrophysics with the Square Kilometre Array
  (AASKA14), \href {http://adsabs.harvard.edu/abs/2015aska.confE..10M} {p.~10}

\bibitem[\protect\citeauthoryear{{Mertens}, {Ghosh}  \& {Koopmans}}{{Mertens}
  et~al.}{2018}]{mertens2018}
{Mertens} F.~G.,  {Ghosh} A.,   {Koopmans} L.~V.~E.,  2018, \mn@doi [\mnras]
  {10.1093/mnras/sty1207}, \href
  {http://adsabs.harvard.edu/abs/2018MNRAS.478.3640M} {478, 3640}

\bibitem[\protect\citeauthoryear{{Mesinger}, {Greig}  \& {Sobacchi}}{{Mesinger}
  et~al.}{2016}]{mesinger2016}
{Mesinger} A.,  {Greig} B.,   {Sobacchi} E.,  2016, \mn@doi [\mnras]
  {10.1093/mnras/stw831}, \href
  {http://adsabs.harvard.edu/abs/2016MNRAS.459.2342M} {459, 2342}

\bibitem[\protect\citeauthoryear{{Mitchell}, {Greenhill}, {Wayth}, {Sault},
  {Lonsdale}, {Cappallo}, {Morales}  \& {Ord}}{{Mitchell}
  et~al.}{2008}]{mitchell2008}
{Mitchell} D.~A.,  {Greenhill} L.~J.,  {Wayth} R.~B.,  {Sault} R.~J.,
  {Lonsdale} C.~J.,  {Cappallo} R.~J.,  {Morales} M.~F.,   {Ord} S.~M.,  2008,
  \mn@doi [IEEE Journal of Selected Topics in Signal Processing]
  {10.1109/JSTSP.2008.2005327}, \href
  {http://adsabs.harvard.edu/abs/2008ISTSP...2..707M} {2, 707}

\bibitem[\protect\citeauthoryear{{Morales} \& {Wyithe}}{{Morales} \&
  {Wyithe}}{2010}]{morales2010rev}
{Morales} M.~F.,  {Wyithe} J.~S.~B.,  2010, \mn@doi [\araa]
  {10.1146/annurev-astro-081309-130936}, \href
  {http://adsabs.harvard.edu/abs/2010ARA%26A..48..127M} {48, 127}

\bibitem[\protect\citeauthoryear{{Mort}, {Dulwich}, {Salvini}, {Adami}  \&
  {Jones}}{{Mort} et~al.}{2010}]{mort2010}
{Mort} B.~J.,  {Dulwich} F.,  {Salvini} S.,  {Adami} K.~Z.,   {Jones} M.~E.,
  2010, in IEEE International Symposium on Phased Array Systems and Technology.
  IEEE, pp 690--694, \mn@doi{10.1109/ARRAY.2010.5613289}

\bibitem[\protect\citeauthoryear{{Offringa} et~al.,}{{Offringa}
  et~al.}{2014}]{offringa2014}
{Offringa} A.~R.,  et~al., 2014, \mn@doi [\mnras] {10.1093/mnras/stu1368},
  \href {http://adsabs.harvard.edu/abs/2014MNRAS.444..606O} {444, 606}

\bibitem[\protect\citeauthoryear{{Parsons}, {Pober}, {Aguirre}, {Carilli},
  {Jacobs}  \& {Moore}}{{Parsons} et~al.}{2012}]{parsons2012}
{Parsons} A.~R.,  {Pober} J.~C.,  {Aguirre} J.~E.,  {Carilli} C.~L.,  {Jacobs}
  D.~C.,   {Moore} D.~F.,  2012, \mn@doi [\apj] {10.1088/0004-637X/756/2/165},
  \href {http://adsabs.harvard.edu/abs/2012ApJ...756..165P} {756, 165}

\bibitem[\protect\citeauthoryear{{Press} \& {Schechter}}{{Press} \&
  {Schechter}}{1974}]{press1974}
{Press} W.~H.,  {Schechter} P.,  1974, \mn@doi [\apj] {10.1086/152650}, \href
  {http://adsabs.harvard.edu/abs/1974ApJ...187..425P} {187, 425}

\bibitem[\protect\citeauthoryear{{Pritchard} \& {Loeb}}{{Pritchard} \&
  {Loeb}}{2010}]{pritchard2010}
{Pritchard} J.~R.,  {Loeb} A.,  2010, \mn@doi [\prd]
  {10.1103/PhysRevD.82.023006}, \href
  {http://adsabs.harvard.edu/abs/2010PhRvD..82b3006P} {82, 023006}

\bibitem[\protect\citeauthoryear{{Remazeilles}, {Dickinson}, {Banday},
  {Bigot-Sazy}  \& {Ghosh}}{{Remazeilles} et~al.}{2015}]{remazeilles2015}
{Remazeilles} M.,  {Dickinson} C.,  {Banday} A.~J.,  {Bigot-Sazy} M.-A.,
  {Ghosh} T.,  2015, \mn@doi [\mnras] {10.1093/mnras/stv1274}, \href
  {http://adsabs.harvard.edu/abs/2015MNRAS.451.4311R} {451, 4311}

\bibitem[\protect\citeauthoryear{Ripley}{Ripley}{1996}]{ripley1996}
Ripley B.~D.,  1996, {Pattern Recognition and Neural Networks}.
Cambridge University Press, \mn@doi{10.1017/CBO9780511812651}

\bibitem[\protect\citeauthoryear{{Rumelhart}, {Hinton}  \&
  {Williams}}{{Rumelhart} et~al.}{1986}]{rumelhart1986}
{Rumelhart} D.~E.,  {Hinton} G.~E.,   {Williams} R.~J.,  1986, \mn@doi [\nat]
  {10.1038/323533a0}, \href {http://adsabs.harvard.edu/abs/1986Natur.323..533R}
  {323, 533}

\bibitem[\protect\citeauthoryear{{Shen}, {George}, {Huerta}  \& {Zhao}}{{Shen}
  et~al.}{2017}]{shen2017}
{Shen} H.,  {George} D.,  {Huerta} E.~A.,   {Zhao} Z.,  2017, preprint, \href
  {http://adsabs.harvard.edu/abs/2017arXiv171109919S} {} (\mn@eprint {arXiv}
  {1711.09919})

\bibitem[\protect\citeauthoryear{{Simonyan} \& {Zisserman}}{{Simonyan} \&
  {Zisserman}}{2014}]{simonyan2014}
{Simonyan} K.,  {Zisserman} A.,  2014, preprint, \href
  {http://adsabs.harvard.edu/abs/2014arXiv1409.1556S} {} (\mn@eprint {arXiv}
  {1409.1556})

\bibitem[\protect\citeauthoryear{{Springenberg}, {Dosovitskiy}, {Brox}  \&
  {Riedmiller}}{{Springenberg} et~al.}{2015}]{springenberg2015}
{Springenberg} J.~T.,  {Dosovitskiy} A.,  {Brox} T.,   {Riedmiller} M.,  2015,
  in International Conference on Learning Representations (ICLR 2015).
  (\mn@eprint {arXiv} {1412.6806})

\bibitem[\protect\citeauthoryear{{Suganuma}, {Ozay}  \& {Okatani}}{{Suganuma}
  et~al.}{2018}]{suganuma2018}
{Suganuma} M.,  {Ozay} M.,   {Okatani} T.,  2018, in Proceedings of the 35th
  International Conference on Machine Learning (ICML 2018). PMLR, p.~4771
  (\mn@eprint {arXiv} {1803.00370})

\bibitem[\protect\citeauthoryear{{Szegedy} et~al.,}{{Szegedy}
  et~al.}{2015}]{szegedy2015}
{Szegedy} C.,  et~al., 2015, in IEEE Conference on Computer Vision and Pattern
  Recognition (CVPR 2015). IEEE, pp~1--9 (\mn@eprint {arXiv} {1409.4842}),
  \mn@doi{10.1109/CVPR.2015.7298594}

\bibitem[\protect\citeauthoryear{{Vafaei Sadr}, {Vos}, {Bassett}, {Hosenie},
  {Oozeer}  \& {Lochner}}{{Vafaei Sadr} et~al.}{2019}]{vafaeiSadr2019}
{Vafaei Sadr} A.,  {Vos} E.~E.,  {Bassett} B.~A.,  {Hosenie} Z.,  {Oozeer} N.,
   {Lochner} M.,  2019, \mnras, \href
  {http://adsabs.harvard.edu/abs/2018arXiv180702701V} {484, 2793}

\bibitem[\protect\citeauthoryear{Vincent, Larochelle, Bengio  \&
  Manzagol}{Vincent et~al.}{2008}]{vincent2008}
Vincent P.,  Larochelle H.,  Bengio Y.,   Manzagol P.-A.,  2008, in Proceedings
  of the 25th International Conference on Machine Learning (ICML 2008). ACM, pp
  1096--1103, \mn@doi{10.1145/1390156.1390294}

\bibitem[\protect\citeauthoryear{Vincent, Larochelle, Lajoie, Bengio  \&
  Manzagol}{Vincent et~al.}{2010}]{vincent2010}
Vincent P.,  Larochelle H.,  Lajoie I.,  Bengio Y.,   Manzagol P.-A.,  2010,
  The Journal of Machine Learning Research, 11, 3371

\bibitem[\protect\citeauthoryear{{Wang}, {Tegmark}, {Santos}  \& {Knox}}{{Wang}
  et~al.}{2006}]{wang2006}
{Wang} X.,  {Tegmark} M.,  {Santos} M.~G.,   {Knox} L.,  2006, \mn@doi [\apj]
  {10.1086/506597}, \href {http://adsabs.harvard.edu/abs/2006ApJ...650..529W}
  {650, 529}

\bibitem[\protect\citeauthoryear{{Wang} et~al.,}{{Wang}
  et~al.}{2010}]{wang2010}
{Wang} J.,  et~al., 2010, \mn@doi [\apj] {10.1088/0004-637X/723/1/620}, \href
  {http://adsabs.harvard.edu/abs/2010ApJ...723..620W} {723, 620}

\bibitem[\protect\citeauthoryear{{Wang} et~al.,}{{Wang}
  et~al.}{2013}]{wang2013}
{Wang} J.,  et~al., 2013, \mn@doi [\apj] {10.1088/0004-637X/763/2/90}, \href
  {http://adsabs.harvard.edu/abs/2013ApJ...763...90W} {763, 90}

\bibitem[\protect\citeauthoryear{Wang, Huang, Wang  \& Wang}{Wang
  et~al.}{2014}]{wang2014}
Wang W.,  Huang Y.,  Wang Y.,   Wang L.,  2014, in IEEE Conference on Computer
  Vision and Pattern Recognition Workshops. IEEE, pp 496--503,
  \mn@doi{10.1109/CVPRW.2014.79}

\bibitem[\protect\citeauthoryear{{Wilman} et~al.,}{{Wilman}
  et~al.}{2008}]{wilman2008}
{Wilman} R.~J.,  et~al., 2008, \mn@doi [\mnras]
  {10.1111/j.1365-2966.2008.13486.x}, \href
  {http://adsabs.harvard.edu/abs/2008MNRAS.388.1335W} {388, 1335}

\bibitem[\protect\citeauthoryear{Xie, Xu  \& Chen}{Xie et~al.}{2012}]{xie2012}
Xie J.,  Xu L.,   Chen E.,  2012, in Proceedings of the 25th International
  Conference on Neural Information Processing Systems (NIPS 2012). Curran
  Associates Inc., USA, pp 341--349

\bibitem[\protect\citeauthoryear{{Zeiler} \& {Fergus}}{{Zeiler} \&
  {Fergus}}{2014}]{zeiler2014}
{Zeiler} M.~D.,  {Fergus} R.,  2014, in Fleet D.,  Pajdla T.,  Schiele B.,
  Tuytelaars T.,  eds, European Conference on Computer Vision (ECCV 2014).
  Springer-Verlag, pp 818--833 (\mn@eprint {arXiv} {1311.2901}),
  \mn@doi{10.1007/978-3-319-10590-1_53}

\makeatother
\end{thebibliography}



\bsp	
\label{lastpage}
\end{document}